\documentclass[12pt,preprint]{aastex}
\newcommand{\be}{\begin{equation}}
\newcommand{\ee}{\end{equation}}
\newcommand{\bea}{\begin{eqnarray}}
\newcommand{\eea}{\end{eqnarray}}
\begin{document}
\title{Simulations of MHD Turbulence in a Strongly Magnetized Medium}
\author{Jungyeon Cho \& Alex Lazarian}
\affil{Department of Astronomy, University of Wisconsin,
    475 N. Charter St.,Madison, WI53706; 
    cho@astro.wisc.edu, lazarian@astro.wisc.edu}

\author{Ethan T. Vishniac}
\affil{Department of Physics and Astronomy, Johns Hopkins University,
Baltimore, MD 21218; ethan@pha.jhu.edu}


\begin{abstract}
We analyze 3D numerical simulations of driven incompressible
magnetohydrodynamic (MHD) turbulence in a periodic box threaded by a
moderately strong external magnetic field.  We sum over nonlinear
interactions within Fourier wavebands and find that the time scale for the
energy cascade is consistent with the Goldreich-Sridhar model of
strong MHD turbulence.  Using higher order longitudinal structure
functions we show that the turbulent motions in the plane perpendicular
to the local mean magnetic field are similar to ordinary hydrodynamic
turbulence while motions parallel to the field are
consistent with a scaling correction which arises from the eddy
anisotropy.
We present the structure tensor describing
velocity statistics of Alfvenic and pseudo-Alfvenic turbulence.  Finally,
we confirm that an imbalance of energy moving up and down 
magnetic field lines leads to a 
slow decay of turbulent motions and speculate that this imbalance
is common in the interstellar medium where injection of energy is
intermittent both in time and space.

\end{abstract}


\section{Introduction}

The interstellar medium (ISM) is complicated and dynamic.
The magnetic field and dynamic pressure ($\rho\,{\bf
v}^2$/2) usually dominate the thermal pressure ($nkT$),
dramatically influencing the star formation rate (see McKee
1999 for a review). There are cosmic rays that provide
pressure and heating as well.

One approach to studying the ISM is to perform time-dependent numerical
simulations to model the ISM, including as many of the
interacting phenomena as practical. Of course, the physics
included in such models must necessarily be highly
simplified, and it is difficult to determine which features
of the final model result from which physical assumptions (or
initial conditions). Our approach is to use simplified
numerical simulations to study the influences of various
physical phenomena in isolation. We want to obtain a
physical feeling for the general effects that each
phenomenon has on the nature of the ISM. In this paper we
will consider the influence of random forces per unit volume
on MHD turbulence in an incompressible medium. It is obvious
that the real ISM is compressible, but we want to separate
the effects of magnetic turbulence from those involving
compression. In later papers we will
include compression for comparison with the present models,
thereby isolating its importance directly.

Historically hydrodynamic turbulence in an incompressible fluid was 
successfully
described by the eddy cascade (Kolmogorov 1941), 
but MHD turbulence was first modeled by 
wave turbulence (Iroshnikov 1963, Kraichnan 1965; hereinafter IK). 
This theory assumes isotropy of the energy cascade
in Fourier space, an assumption which has attracted severe criticism
(Montgomery \& Turner 1981; Shebalin et al. 1983; 
Montgomery \& Matthaeus 1995;
Sridhar \& Goldreich 1994).
Indeed, the magnetic field defines a local symmetry axis since 
it is easy to mix
field lines in directions perpendicular to the
local ${\bf B}$ and much more difficult to bend them.
   The idea of an anisotropic (perpendicular) cascade has been incorporated 
   into
   the framework of the reduced MHD approximation
  (Strauss 1976; Rosenbluth 1976; Montgomery 1982; Zank \& Matthaeus 1992;
  Bhattacharjee, Ng, \& Spangler 1998).

In a turbulent medium, the kinetic energy associated with large scale motions
is greater than that of small scales.
However, the strength of the local mean
magnetic field is almost the same on
all scales.
Therefore, it becomes relatively difficult to bend
magnetic field lines as we consider smaller scales,
leading to more pronounced anisotropy.
A self-consistent model of MHD turbulence which incorporates
this concept of scale dependent anisotropy was introduced
by Goldreich \& Sridhar (1995) (henceforth GS95).

Within the GS95 theory the energy cascade becomes anisotropic
as a consequence of the resonant conditions for 3-wave
interactions.
A strict application of the resonant 3-wave interaction
conditions gives an energy cascade which is purely in the direction 
perpendicular to the external field. However, it is intuitively
clear that the increase in $k_{\perp}$ must at some point start
affecting $k_{\|}$.

The cornerstone of the GS95 theory is
the concept of a `critically balanced' cascade, where
$k_{\parallel}V_A \sim k_{\perp}v_l$,
where $k_{\perp}$ and $k_{\|}$ are wave numbers perpendicular
and parallel to the background field an $v_l$ is the r.m.s. speed of
turbulence at the scale $l$.
In this model, the Alfv\'en rate ($k_{\parallel}V_A$)
is equal to the eddy turnover rate ($k_{\perp}v_l$).
Using this concept, Goldreich \& Sridhar showed that the energy cascade is not
strictly perpendicular to the background field, but is relaxed so that
$k_{\|}\propto k_{\perp}^{2/3}$.
Their model predicts that the one-dimensional energy spectrum 
is Kolmogorov-type if expressed in terms of perpendicular wavenumbers, i.e.
$E(k_{\perp})\propto k_{\perp}^{-5/3}$.

Numerical simulations by Cho \& Vishniac (2000a, henceforth CV00) and 
Maron \& Goldreich (2001,
hereinafter MG01)
have mostly supported the GS95 model and helped to extend it. 
Both analyses stressed the point 
that scale dependent anisotropy can be measured only in 
local coordinate frames which are aligned with the locally averaged
magnetic field direction.  Cho \& Vishniac (2000a) calculated
the structure functions of velocity and magnetic field in the
local frames, and found that the contours of the structure functions do 
show scale dependent
anisotropy, consistent with the predictions of the GS95 model.
In their calculation, the strength of the uniform background magnetic field is
roughly the same as the the r.m.s. velocity.
MG01 tested the GS95 model for a much stronger uniform background
field and also obtained results supporting the GS95 model, but 
they produced $E(K)\propto k^{-3/2}$.
They also calculated time scales of turbulence, interactions between
pseudo- and shear-Alfvenic modes, growth of imbalance, and intermittency.
     Other related recent numerical simulations include 
     Matthaeus et al. (1998),
     M\"uller \& Biskamp (2000), and Milano et al. (2001).

These studies left a number of unresolved issues, including the exact 
scaling relations,
the comparison of intermittency in MHD and in hydrodynamic turbulence,
and the time scale of turbulence decay. Moreover, for many practical
applications a more quantitative description of MHD turbulence
statistics is necessary.
These are vital for understanding various astrophysical
processes, including star formation (McKee 1999), cosmic ray propagation
(K\'ota \& Jokipii 2000), and
magnetic reconnection (Lazarian \& Vishniac 1999).

In this paper, we further investigate implications of the GS95 model.
In \S2, we explain our numerical method. In \S3, we further elucidate
the scaling relation implied by the GS95 model.
In particular, we discuss the time scale, velocity scaling relations,
and intermittency.
In \S4, we derive the correlation tensor and discuss some astrophysical
applications.  While the GS95 model predicts the
MHD turbulence decays in just one eddy turnover time,
in \S5, we show that the decay time scale increases when the cascade
is unbalanced and discuss some consequences of this fact.
In \S6, we briefly discuss the implications of this work. In
\S7 we give a summary and our conclusions.  As before, we consider
the case where the uniform background magnetic field energy density is
comparable to the turbulent energy density.


\section{Method}

\subsection{Numerical Method}
We have calculated the time evolution of incompressible magnetic turbulence
subject to a random driving force per unit mass.
We have adopted a pseudospectral code to solve the
incompressible MHD equations in a periodic box of size $2\pi$:
\begin{equation}
\frac{\partial {\bf v} }{\partial t} = (\nabla \times {\bf v}) \times {\bf v}
      -(\nabla \times {\bf B})
        \times {\bf B} + \nu \nabla^{2} {\bf v} + {\bf f} + \nabla P' ,
        \label{veq}
\end{equation}
\begin{equation}
\frac{\partial {\bf B}}{\partial t}=
     \nabla \times ({\bf v} \times{\bf B}) + \eta \nabla^{2} {\bf B} ,
     \label{beq}
\end{equation}
\be
      \nabla \cdot {\bf v} =\nabla \cdot {\bf B}= 0,
\ee
where $\bf{f}$ is a random driving force,
$P'\equiv P/\rho + {\bf v}\cdot {\bf v}/2$, ${\bf v}$ is the velocity,
and ${\bf B}$ is magnetic field divided by $(4\pi \rho)^{1/2}$.
In this representation, ${\bf v}$ can be viewed as the velocity 
measured in units of the r.m.s. velocity, v,
of the system and ${\bf B}$ as the Alfven speed in the same units.
The time $t$ is in units of the large eddy turnover time ($\sim L/v$) and
the length in units of $L$, the inverse wavenumber of the fundamental
box mode.
In this system of units, the viscosity $\nu$ and magnetic diffusivity $\eta$
are the inverse of the kinetic and magnetic Reynolds numbers respectively.
The magnetic field consists of the uniform background field and a
fluctuating field: ${\bf B}= {\bf B}_0 + {\bf b}$.
We use 21 forcing components with $2\leq k \leq \sqrt{12}$, where
wavenumber $k$ is in units of $L^{-1}$.
Each forcing component has correlation time of one.
The peak of energy injection occurs at $k\approx 2.5 $.
The amplitudes of the forcing components are tuned to ensure $v \approx 1$
We use exactly the same forcing terms
for all simulations.
The Alfv\'en velocity of
the uniform background field, $B_0$, is set to 1.
We consider only cases where viscosity is
equal to magnetic diffusivity:
\be
  \nu = \eta.
\ee
In pseudo spectral methods, the temporal evolution of
equations (\ref{veq}) and (\ref{beq}) are followed in Fourier space.
To obtain the Fourier components of nonlinear terms, we first calculate
them in real space, and transform back into Fourier space.
The average kinetic helicity in these simulations is not zero. However,
previous tests have shown that our results are insensitive to the value of the 
kinetic helicity.
In incompressible fluid, $P'$ is not an independent variable.
We use an appropriate projection operator to calculate
$\nabla P'$ term in
{}Fourier space and also to enforce divergence-free condition
($\nabla \cdot {\bf v} =\nabla \cdot {\bf B}= 0$).
We use up to $256^3$ collocation points.
We use an integration factor technique for kinetic and magnetic dissipation terms
and a leap-frog method for nonlinear terms.
We eliminate the $2\Delta t$ oscillation of the leap-frog method by using
an appropriate average.
At $t=0$, the magnetic field has only its uniform component
and the velocity field is restricted to the range
$2\leq k \leq 4$ in wavevector space.

Hyperviscosity and hyperdiffusivity are
used for the dissipation terms (see Table 1).
The power of hyperviscosity
is set to 8, so that the dissipation term in the above equation
is replaced with
\be
 -\nu_8 (\nabla^2)^8 {\bf v},
\ee
where $\nu_8$ is determined from the condition $\nu_h (N/2)^{2h} \Delta t
\approx 0.5$ (see Borue and Orszag 1996)
Here $\Delta t$ is the time step and $N$ is the number
of grid points in each direction.
The same expression is used for the magnetic dissipation term.
We list parameters used for the simulations in Table 1.
We use the notation XY-$B_0$Z,
where X = 256, 144 refers to the number of grid points in each spatial
direction; Y = H refers to hyperviscosity;
Z=1 refers to the strength of the external magnetic field.

Diagnostics for our code can be found in Cho and Vishniac (2000b).
 {}For example, our code conserves total energy
   very well in simulations with $\nu=\eta=0$ 
   and the average energy input (=${\bf f}\cdot {\bf v}$)
   is almost exactly the same as the sum of magnetic and viscous
dissipation in simulations with nonzero $\nu$ and $\eta$.
The runs 256H-$B_0$1 and 144H-$B_0$1 are exactly the same as the runs
256H-$B_0$1 and REF2 in CV00.
The energy spectra as a function of time for these runs can be found in that paper.


\subsection{Defining the Local Frame}
   The GS95  model deals with
   strong MHD turbulence and should be distinguished
   from theories that deal with weak MHD turbulence (e.g. 
Sridhar \& Goldreich 1994;
   Ng \& Bhattacharjee 1996; Galtier et al. 2000; see also 
Goldreich \& Sridhar 1997).  
In strong MHD turbulence
eddy-like motions mix up magnetic field lines perpendicular to the local
direction of magnetic field. Thus, as in the case of hydrodynamic turbulence,
the correlation time for coherent structures
is comparable to the inverse of $k_{\perp}v_k$ for any 
scale $k_{\perp}^{-1}$. These mixing motions are strongly coupled
to wave-like motions
with a correlation time  $(k_{\|}V_A)^{-1}$.
The GS95 model is based on the concept of a `critical balance' between
these time scales, that is 
   $k_{\parallel}V_A \sim k_{\perp}v_k$. . This results in
a scale dependent
   anisotropy, $k_{\|} \propto k_{\perp}^{2/3}$ so that the eddies
are increasingly elongated on smaller scales.

The turbulent magnetic field changes its direction in the {\it global}
system of reference.
It is important that the mixing motions are available only in the
direction perpendicular to the {\it local} direction of magnetic 
field. Thus the theory must be formulated using the system of
reference aligned with the local magnetic field. CV01 discusses
in detail one way of defining this system given numerical data.

 Figure 1 is a schematic representation of the GS95 model.\footnote{
      Note that Figure 1 is {\it local frame representation} of the 
      GS95 model. See below for more details.}
In Fourier space, the energy injected on
large scales excites large scale Fourier components of the magnetic field
(the dark region at the center in Fig. 1a).
The external magnetic field makes the subsequent energy cascade to
small scales anisotropic: it occurs in the directions perpendicular to the
mean external field.
The GS95 model states that
most of the energy is confined to the region 
$k_{\|}=\pm k_{\perp}^{2/3}$,
and as the energy cascades to larger values of $k_{\perp}$
the energy of Fourier components between $k_{\perp}$ and $k_{\perp}+1$ 
decreases as
$E(k_{\perp})\propto k_{\perp}^{-5/3}$.

\begin{figure}
\plottwo{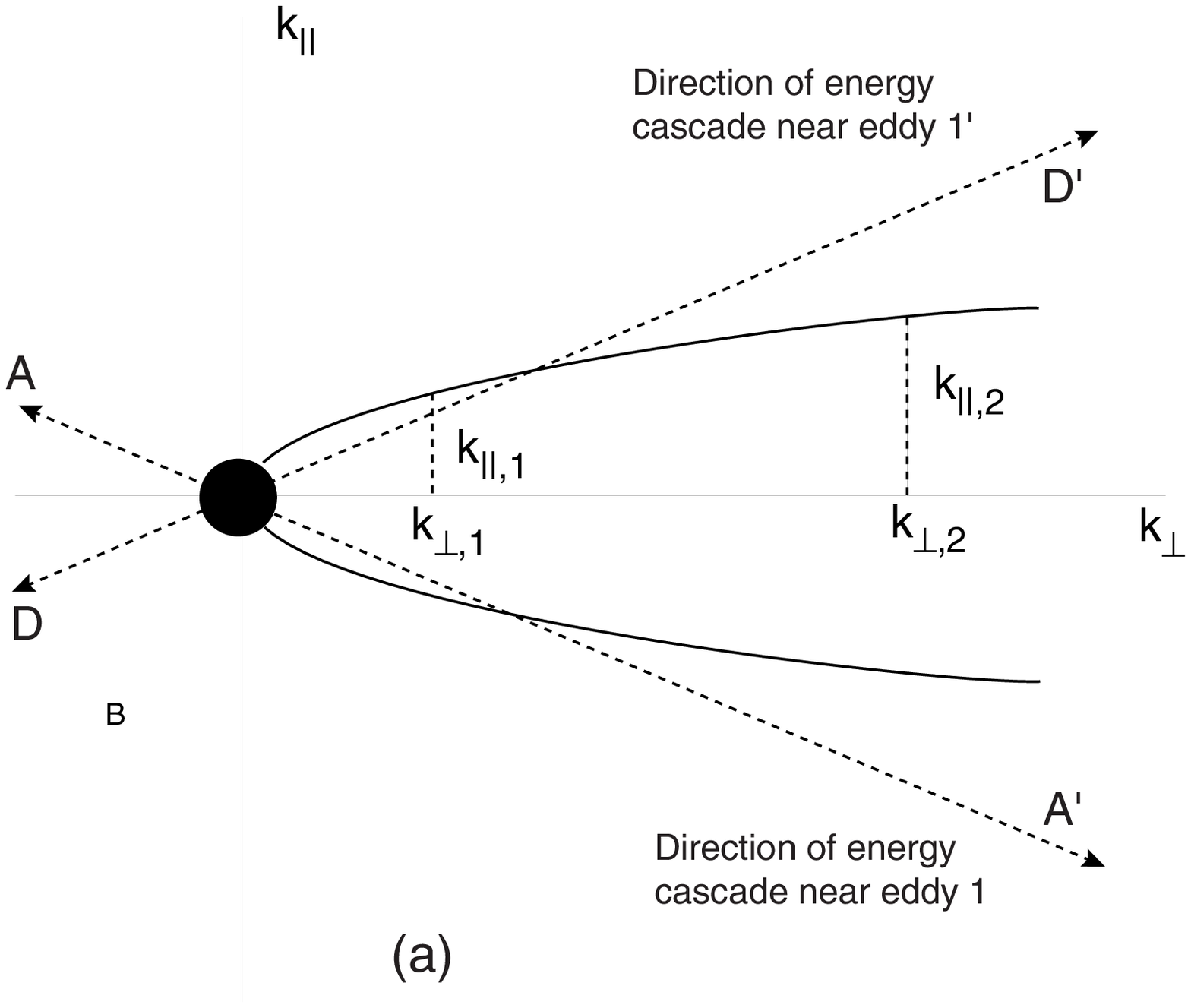}{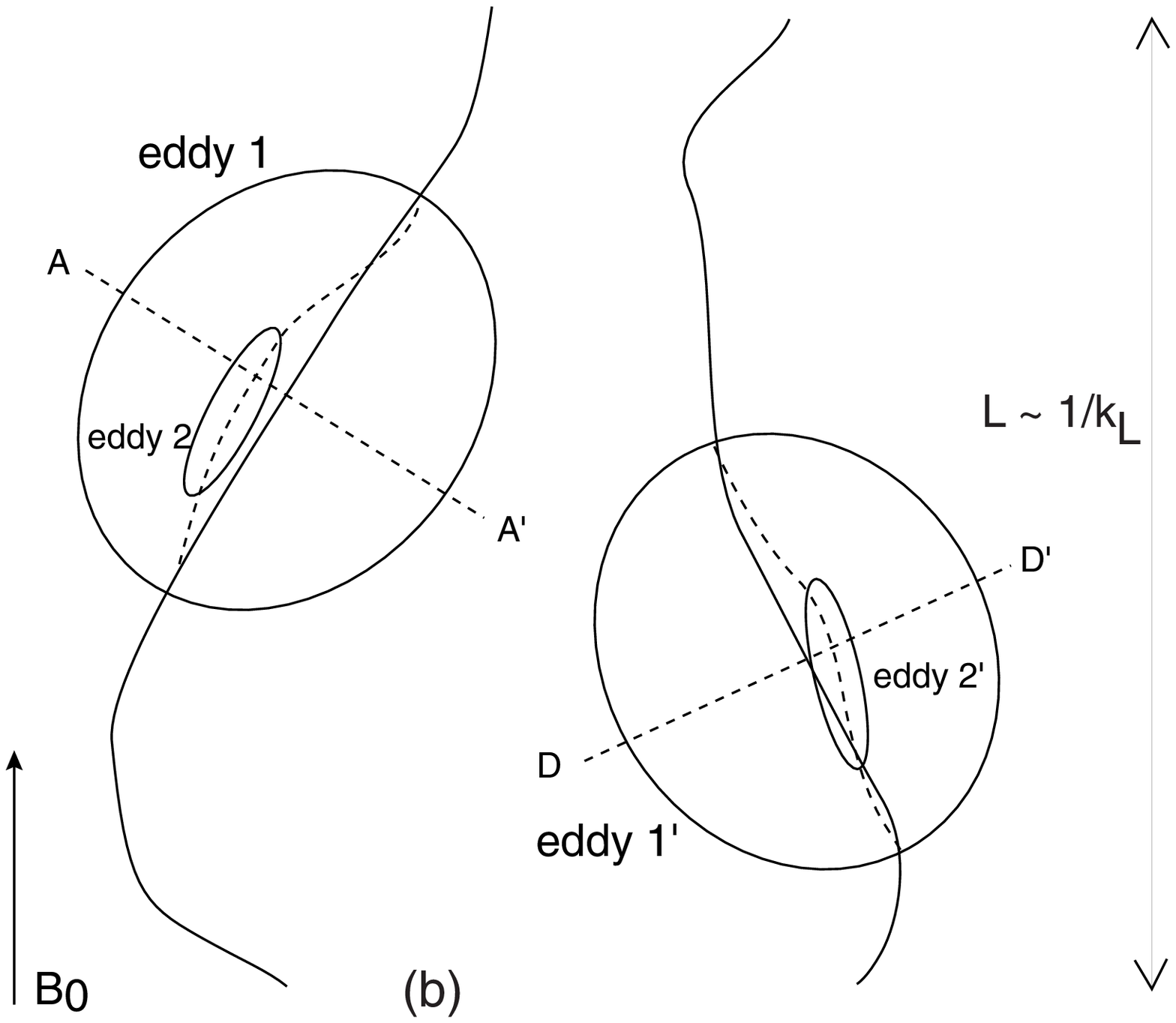}
\caption{ (a) Fourier space structure. (b) Real space structure.
    Large eddies (eddy $1$ or $1^{\prime}$)) have similar semi-major axis 
    ($\sim 1/k_{\|,1}$) and
    semi-minor axis ($\sim 1/k_{\perp,1}$).
    Therefore, they are almost isotropic.
    Smaller eddies (eddy $2$ or $2^{\prime}$)
    have relatively larger semi-major axis ($\sim 1/k_{\|,2}$)
    to semi-minor axis ($\sim 1/k_{\perp,2}$) ratio.
    Therefore, they are relatively more elongated.
    Energy cascades in the directions perpendicular to large-scale
    magnetic field lines (e.g. direction  $AA^{\prime}$ or $DD^{\prime}$
    in both figure (a) and
    (b)). This effect obscures scale-dependent anisotropy
     (e.g. $k_{\|}\propto k_{\perp}^{2/3}$ in GS95 model)
    when we perform Fourier analysis in the global frame.
   In Figure (b), the solid curves represent the wandering of 
   magnetic field lines by the large scale magnetic fields.
   The solid curves can define the directions of the local mean magnetic
   field line for eddy $1$ or $1^{\prime}$.
   Similarly, the dashed curves can define 
   the directions of the local mean magnetic
   field line for the eddy $2$ or $2^{\prime}$.
   }

\end{figure}


As illustrated in Fig. 1b, eddies are not aligned along the mean field
${\bf B}_0$. Instead, they are aligned along the {\it local mean} field lines.
The {\it local mean} magnetic field defines the physically relevant 
background for the eddy dynamics, and is determined by the Fourier components 
whose wavenumbers are
a bit less than the characteristic wavenumber of the eddy.
In practice, it can be obtained by averaging the magnetic field in 
the vicinity of the eddy
over a volume slightly (say, for example, 2$\times$) 
larger than the size of the eddy (see CV00 for details).
The solid curves in Fig. 1b represent this kind of locally defined
mean, formed by all magnetic Fourier components whose scales are
a bit larger than eddy 1 (or 1$^{\prime}$).
     The characteristic scale of this wandering is $L\sim 1/k_L$, the energy
     injection scale, because eddy 1 (or 1$^{\prime}$) is only slightly
     smaller than the energy injection scale.
     This large-scale wandering is smooth but
dominates over smaller scale effects because the
the magnetic energy is concentrated on larger scales.
Wandering by smaller scale magnetic fields is weaker and
causes smaller deviations from the large-scale wandering.
     We depict the additional wandering caused by scales a bit larger than 
     the eddy 2 as a dashed
     curve in Figure 1b.  
    For the eddy 1, the solid curve defines the local mean magnetic
     field and, for the eddy 2 the dashed curve.

The GS95
model is dominated by local dynamics, that is, in this model 
disturbances lose their
coherence when propagating over a single wavelength.  To the extent that
the dynamics are local it is obvious that the only relevant magnetic field 
is the local mean field.  As an example, consider eddies 1 and 2 in Figure 1b again.
{}For eddy 1, the solid curve 
can be regarded as a local mean field line and the energy cascade will take
place perpendicular to this field, along the direction $AA^{\prime}$.
Smaller eddies, like eddy 2 sees the dashed curve as its local
background magnetic field with a slightly different direction for the
energy cascade.  Since the difference between the two curves is small, the 
direction of the energy cascade differs only slightly as a function of scale.
Basically, energy cascades along $AA^{\prime}$ in the region near 
the eddy 1.  Similarly, energy cascades along $DD^{\prime}$ in the region near the 
eddy 1$^{\prime}$.  We are left with an energy cascade which differs both as
a function of scale and as a function of location.  However, the dynamics
are {\it not} entirely local, in the sense that disturbances do propagate
along field lines without retaining phase coherence.  Consequently, perpendicular
motions in eddy 1 can lead to similar motions in eddy 1$^{\prime}$, even though
$AA^{\prime}$ and $DD^{\prime}$ are not parallel vectors.  This is one of the
key features of the GS95 model, and carries with it the implication is that dynamic 
variables need to be evaluated in terms of the local mean field direction, rather
than a global coordinate system.  Conversely, the ability to generate a more meaningful
description in terms of a local, and scale-dependent, coordinate system can be
taken as an indirect confirmation of some features of the GS95 model.

In summary, when we want to describe the scaling of eddy shapes, we should
correctly identify the direction of the local mean magnetic field.
When we talk about anisotropy, we talk about anisotropy
with respect to local mean magnetic field lines.
Because of this, it is necessary to introduce a `local' frame in which
the direction of the local mean magnetic field lines is taken as
the parallel direction.
When we consider the GS95 picture
(i.e. $k_{\|}\propto k_{\perp}^{2/3}$) in Fourier space, we are considering
the {\it local} frame in real space, and vice versa.
When we describe turbulence with respect to the {\it global} frame, which is
fixed in real space, the corresponding Fourier space structure no longer
shows the GS95 picture.  Instead, we have a relation close to 
$k_{\|}\propto k_{\perp}$.
This is because when energy cascades along  $AA^{\prime}$, $DD^{\prime}$, or
some intermediate directions
in real space (Fig 1b),
it cascades along the directions between $AA^{\prime}$ and
$DD^{\prime}$ in Fourier space (Fig 1a),
which implies that, when we perform the Fourier transform with
respect to the fixed global frame\footnote{Many astrophysical observations,
for instance, interferometric observations of turbulent HI (Lazarian 1995),
provide the statistics measured in the global frame.}, 
we will get $k_{\|}\propto k_{\perp}$.
The true scaling relation is eclipsed by the wandering of large scale
magnetic field lines.

It is very important to identify the {\it local} frame.
In this paper, when we calculate decay time scale, intermittency, and
the correlation tensor, we always refer to the local frame.


\section{Scaling Relations}


\subsection{Time Scale of Motions}

One of the basic questions in the theory of MHD turbulence is the slope of the
one-dimensional energy spectra. As we have seen, GS95
obtained a spectral index of $-5/3$. In the numerical simulations of 
CV00 the spectral index is close to $-5/3$, while
it is very close to $-3/2$ in MG01. 
The IK theory predicts a $k^{-3/2}$ scaling, although the other features of
this model are definitely inconsistent with all the numerical
evidence.  MG01 attributed their result to the appearance
of strong intermittency in their simulations.  
We note that the inertial range of the solar wind shows a spectral
index of $-1.7$ (Leamon et al. 1998; see also Matthaeus \& Goldstein 1982), 
but this number should be considered
cautiously. The physics of the solar
wind is undoubtedly more complicated than the simulations described
here.

Can we test which scaling is correct? The cascade time as function of
scale presents us with an interesting constraint.

IK theory and GS95 model predict different
scalings for turbulent cascade time scale ($t_{cas}$).
In both theories, $t_{cas}$ can be determined by the
scale-independence of the cascade:
\be
  v^2_k /t_{cas} = \mbox{constant}.
\ee
Since $v^2_k$ is proportional to $kE(k)$,
we have
\be
  t_{cas, IK} \propto k^{-1/2}, ~~~~ t_{cas, GS} \propto k^{-2/3}
\ee
for IK theory and GS95 model respectively.
This result is also useful for certain intermittency theories
(see \S3.3).  MG01 studied the cascade time scale using three different
methods and obtained
slopes comparable to $-2/3$ (i.e. $t_{cas}\propto k_{\perp}^{-2/3}$)
in two methods and $-1/2$ in the other method.

Here we consider a different method of evaluating $t_{cas}$.
The purpose of our calculation is to test MG01's result
using another numerical method and demonstrate
the effects of large scale fluid motions on the calculation of $t_{cas}$.

Symbolically, we can rewrite the MHD equations as follows:
\bea
\dot{\bf v}_{\bf k} = N^v_{\bf k},\\
\dot{\bf b}_{\bf k} = N^b_{\bf k},
\eea
where $N^v$ and $N^b$ represent
the nonlinear terms. We have ignored the dissipation terms.
Naively speaking, we might obtain the time scale
by dividing $|{\bf v}_{\bf k}|$ by $|N^v_{\bf k}|$.
However, this gives $t_{cas} \propto k^{-1}$, where the exponent is
almost exactly $-1$.  This is not actually a measure of the
cascade time.  We note that CV00 obtained
a similarly misleading relation for the cascade time, and
attributed it to the effect of large scale translational motions.
Although they used a different method to calculate the cascade time,
the same argument applies here.  If we consider
the interaction between a small eddy and a large scale 
(translational) fluid motion,
then the translation
can be removed by a Galilean transformation, and
there is no associated energy cascade.
However, the phase of the Fourier components that represent the small eddy
is affected by the large scale translational motion, and
changes at a rate {\it kV}, where $V$ is the large scale velocity.
The corresponding nonlinear term has a magnitude of
$|N_k| \sim |{\bf v}_k|kV$, which accounts for the 
(misleading) relation $t_{cas}\sim k^{-1}$.
The cascade time
as a function of wavenumber can be evaluated directly from our simulations,
but only after we filter out translational motions arising from eddies
much larger than the scale under consideration.

We correct for the presence of large scale motions by restricting the
evaluation of the nonlinear terms to contributions coming from the
interactions between the mode at $k$ and
other modes within the range\footnote{This assumes some sort of locality
which may not be exact in the presence of strong intermittency.}
of $k/2$ and $2k$.
In doing this, we retain the uniform magnetic 
component $B_0$.
We show the result in Figure 2.
Our result supports the GS95 model: $t_{cas}\propto k^{-2/3}$.
In comparison with MG01 we obtained this result using a different 
method and for a different kinetic/magnetic energy ratio.

In the GS95 model, $t_{cas}$ is determined by the relation
$t_{cas}\sim l_{\perp}/v_{l\perp}$.  This means that
the cascade time scale is virtually synonymous with the eddy turnover time, which
is also true for hydrodynamic turbulence.
It is obvious that the cascade time determines the decay time scale of
turbulence.
As a consequence, the GS95 model implies that MHD turbulence decays as fast as
hydrodynamic turbulence (say, $t_{decay}$= a few eddy turnover times).
Note that no matter how strong the external field is, strong MHD turbulence
decays within a few eddy turn over times.  We discuss the implications of this result,
and some limitations,  in \S5.

These results support the original GS95 theory.
However, we are not in a position to directly confront the results of
MG01.  Our simulations differ from theirs
in many ways, including the shape of the computational
box, the range of length scales, and the strength of the uniform background field.
We shall address those issues elsewhere.
In \S3.3, we will discuss what the study of intermittency
implies about the slope of energy spectra.

\begin{figure}
\plottwo{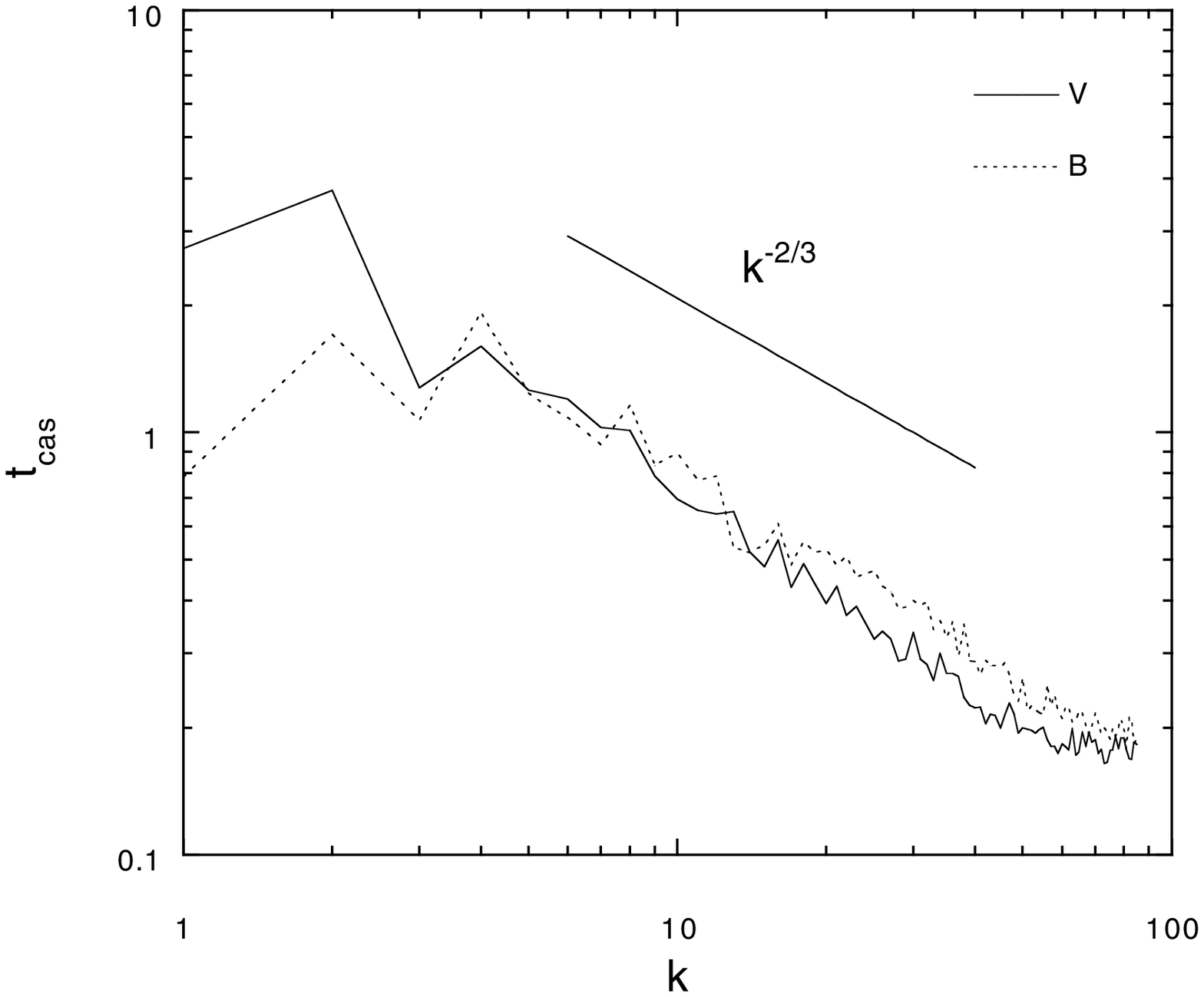}{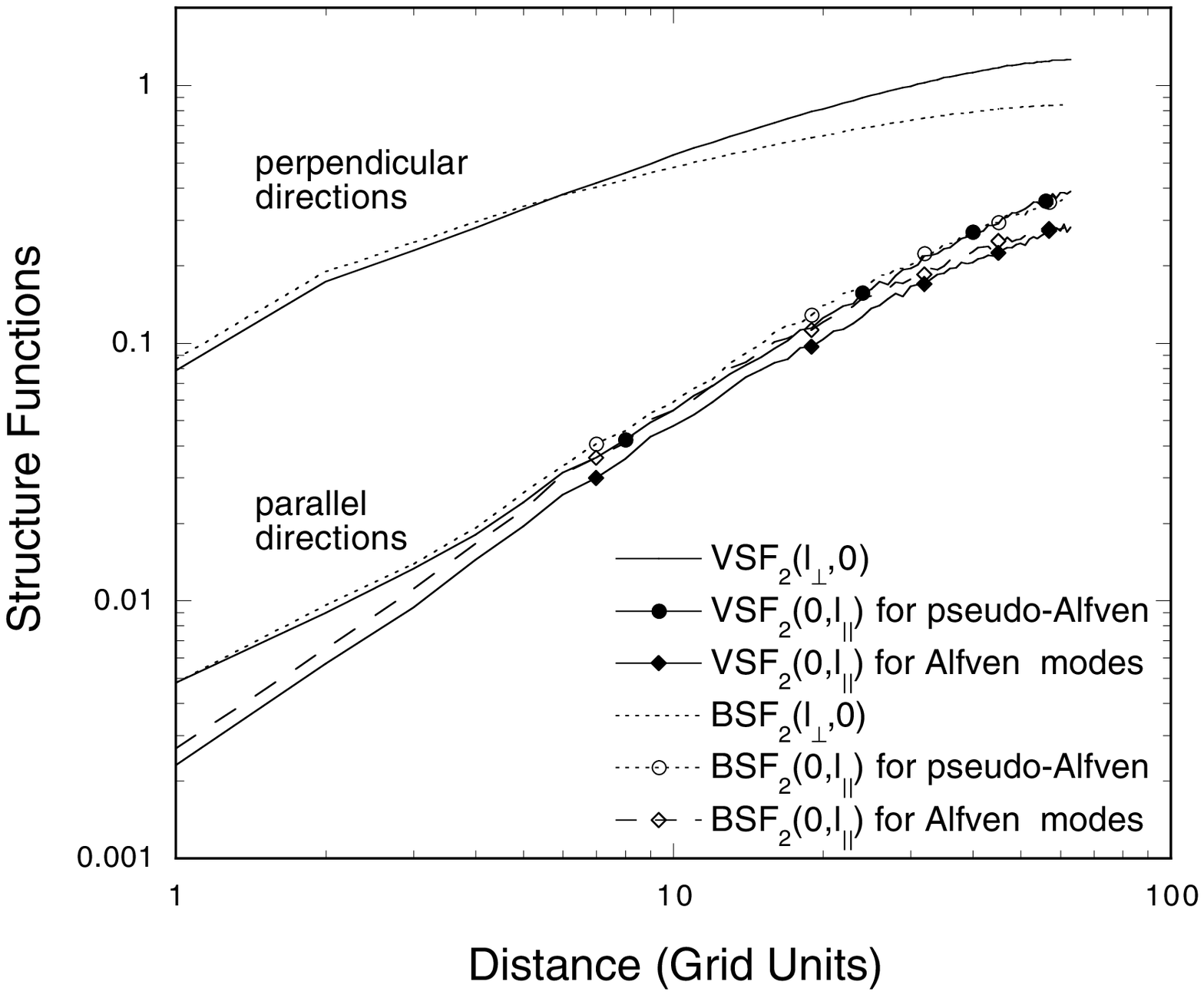}
\caption{   Cascade time scale. The GS95 model predicts
            $t_{cas} \propto k^{-2/3}$,
  while the IK theory predicts $t_{cas} \propto k^{-1/2}$.
  Our result supports the GS95 model. Run 256H-$B_0$1.}  
\caption{
  Second-order structure functions. Across local mean
  magnetic field lines,
  the second-order structure functions follow $r^{2/3}$. Along the
  local mean magnetic field lines, they follow $r^{1}$.
  For pseudo-Alfven modes this defines the scaling of motions parallel to
  local mean magnetic field lines.
  $VSF_2$ and $BSF_2$ denote the second order structure functions for
  velocity and magnetic field respectively.
   Run 256H-$B_0$1.}    
\end{figure}


\subsection{Velocity Scaling}
In the GS95 model, $v_{\perp}^2$ is proportional to
$\sim k_{\perp} E(k_{\perp})$, where
$E(k_{\perp})$ is the one-dimensional energy spectrum.
Since $E(k_{\perp})\propto k_{\perp}^{-5/3}$,
we have $v_{\perp}^2 \propto l_{\perp}^{2/3}$ or
$SF_2(l_{\perp},0) \propto l_{\perp}^{2/3}$,
where $SF_2$ is the second order structure function:
\be
SF_2(l_{\perp},l_{\|})= < \left[ {\bf v}({\bf x}+{\bf l})-
                                      {\bf v}({\bf x})  \right]^2 >,
\ee
where the angle brackets denote the spatial average over {\bf x} and
${\bf l}= l_{\perp} {\bf e}_{\perp} +  l_{\|} {\bf e}_{\|}$. 
The vectors ${\bf e}_{\perp}$ and ${\bf e}_{\|}$ are unit vectors 
perpendicular and parallel to ${\bf B}_L$ respectively.
The vector ${\bf B}_L$ denotes the local mean magnetic field.

What can we say about the velocity scaling parallel
to ${\bf B}_{L}$? We can consider two different quantities.
First, we can consider the scaling
of Alfven components in the direction parallel to  ${\bf B}_{L}$.
Second, we can also consider the scaling of pseudo-Alfvenic components
along the direction of ${\bf B}_{L}$.      
      The pseudo-Alfvenic components are incompressible limit of 
      slow magnetosonic waves. While they have the same 
       dispersion relations as the Alfven modes, the velocities $v_l$ of
     polarization are completely different: the directions lie
    in the plane determined by ${\bf B}_0$ and ${\bf k}$.
Note that the Alfven modes have velocities perpendicular to
      the plane determined by ${\bf B}_0$ and ${\bf k}$.

There are several ways to derive the scaling relation for Alfvenic 
turbulence from the GS95 model. 
First, suppose that the second order structure function along local
${\bf B}_L$ follows a power law:
$SF_2(0,l_{\|})\propto l_{\|}^m$.
When we equate $SF_2(0,l_{\|})$ and  $SF_2(l_{\perp},0)$,
we should retrieve the GS95 scaling relation,  
$ l_{\|}\propto  l_{\perp}^{2/3}$
(see MG01).
We conclude that $m=1$ and $SF_2(0,l_{\|})\propto l_{\|}$.
     Since $k_{\|} E^v (k_{\|}) ~(\propto v_{\|}^2) \propto SF_2(l_{\|},0)$, 
     we have 
    $E^v (k_{\|}) \propto k_{\|}^{-2}$.
Alternatively, we can write
\be
  E_3^v(k_{\perp},k_{\|}) \propto k_{\perp}^{-10/3} g(k_{\|}/k_{\perp}^{2/3}),
  \label{eq10}
\ee
where
$E_3^v(k_{\perp},k_{\|})$ is the 3-dimensional energy spectrum
and $g$ is a function which describes distribution of energy along
$k_{\|}$ direction in Fourier space. We will give a reasonable fit to its
functional form in the next section.

We plot our results in Figure 3, in which we observe that
$SF_2(l_{\perp},0)$ (across ${\bf B}_L$) $\propto l_{\perp}^{2/3}$ and
$SF_2(0, l_{\|})$ (parallel to ${\bf B}_L$) $\propto l_{\|}$.
The velocity field follows these relations quite well, while the
magnetic field follows a slightly different relation across ${\bf B}_L$.
Both Alfvenic
and pseudo-Alfvenic components follow 
similar scalings in the directions parallel to ${\bf B}_L$
In section~4 we shall also show that the 3D spectrum of pseudo-Alfven motions
has a form similar to (\ref{eq10}). On this basis we conclude
that the scaling $E^v(k_{\|})\propto k_{\|}^{-2}$ also applies
to pseudo-Alfven motions, where the velocities are mostly 
parallel to the local mean
magnetic field ${\bf B_L}$. The corresponding velocities
$ v_{\|}^2 \propto k_{\|} E^v(k_{\|})$ which means that
 $ v_{\|}^2 \propto k_{\|}^{-1} \propto l_{\|}$. This result
is important for many problems, including dust transport
(Yan et al 2001).
Note that the energy spectrum is steeper when expressed as a function of the
parallel direction.

In this subsection we extended the GS95 model for the parallel motions 
and pseudo-Alfven 
modes and confirmed it through numerical simulations.


\subsection{Intermittency}

MG01 studied the intermittency of dissipation structures
in MHD turbulence using the fourth order moments of the Elsasser fields and
the gradients of the fields.  Their simulations show strong
intermittent structures.
We use a different, but complementary,  method to study intermittency, 
based on the higher order longitudinal structure functions.
Our result is that by this measure the intermittency of velocity field
in MHD turbulence across local
magnetic field lines is as strong, but not stronger, than in 
hydrodynamic turbulence.

In fully developed hydrodynamic turbulence, the (longitudinal)
velocity structure functions
$S_p=< ( [ {\bf v}({\bf x}+ {\bf r}) -
      {\bf v}({\bf x})]\cdot \hat{\bf r} )^p>
\equiv < \delta v_L^p({\bf r}) >$ are
expected to scale as $r^{\zeta_p}$.
For example, the classical Kolmogorov phenomenology (K41) predicts
$\zeta_p =p/3$.
The (exact) result for p=3 is the well-known 4/5-relation:
$ < \delta v_L^p({\bf r}) > =-(4/5)\epsilon r$, where $\epsilon$
is the energy injection rate (or, energy dissipation rate).
On the other hand, She \& Leveque (1994, hereinafter S-L) 
proposed a different
scaling relation: $\zeta_p^{SL}=p/9+2[1-(2/3)^{p/3}]$.
Note that She-Leveque model also implies $\zeta_3 =1$.

     So far in MHD turbulence, to the best of our best knowledge, 
there is no rigorous intermittency theory which takes into account
     scale-dependent anisotropy. Therefore, we will use 
     an intermittency model based on an extension of a hydrodynamic model.
Politano \& Pouquet (1995) have developed an MHD version of She-Leveque model:
\be
\zeta_p^{PP}=\frac{p}{g}(1-x)+C \left(1-(1-x/C)^{p/g}\right),
\label{mhdint}
\ee
where $C$ is the co-dimension of the dissipative structure,
$g$ is related to the scaling $v_l \sim l^{1/g}$,
and $x$ can be interpreted as the exponent of the cascade time
$t_{cas}\propto l^{x}$.
(In fact,
$g$ is related to the scaling of Elsasser variable z: $z_l\sim l^{1/g}$.)
In the framework of the IK theory, where $g=4$, $x=1/2$, and $C=1$ when
the dissipation structures are sheet-like, their model of intermittency becomes
$\zeta_p^{IK}=p/8+1-(1/2)^{p/4}$.  On the other hand,
M\"uller \& Biskamp (2000) performed numerical simulations on decaying
isotropic MHD turbulence and obtained
Kolmogorov-like scaling ($E(k)\sim k^{-5/3}$ and $t\sim l^{2/3}$) and
sheet-like dissipation structures, which 
implies $g=3$, $C=1$, and $x=2/3$.
{}From equation (\ref{mhdint}), they proposed that
\be
\zeta_p^{MB}=p/9+1-(1/3)^{p/3}.
\ee

How does anisotropy change intermittency?
We have determined the scaling exponents numerically, working
in the local frame.
We performed a simulation with a grid of $144^3$ collocation points
and integrated the MHD equations from t=75 to 120.
We calculated the higher order velocity structure functions for 75 evenly
spaced snapshots.  We average over 5 consecutive values since the
correlation time of the turbulence corresponds to 5 snapshots.
We calculated the scaling exponents from these averaged structure functions.
We obtained a total of 15 (=75/5) such structure functions
and scaling exponents.
We believe these 15 data sets are mutually independent.
We plot the result in Figure 4a.
The filled circles represent 
the scaling exponents of longitudinal
velocity structure functions in directions {\it perpendicular} to
the local mean magnetic field.
It is surprising that the scaling exponents are close the
original (i.e. hydrodynamic) S-L model.
This raises an interesting  question.
In our simulations, we clearly observe that $t_{cas}\propto l^{2/3}$ and
$E(k) \propto k^{-5/3}$.
It is evident that MHD turbulence has sheet-like dissipation structures
(Politano, Pouquet \& Sulem 1995).
Therefore, the parameters for our simulations should be
the same as those of M\"uller \& Biskamp's (i.e. $g=3$, $C=1$, and $x=2/3$)
rather than suggesting $C=2$.
We believe that this difference stems from the different
simulation settings: their turbulence is isotropic and ours is
anisotropic.  In fact, we expect the the small scale behavior of MHD turbulence
should not depend on whether or not the largest scale fields are uniform or
have the same scale of organization as the largest turbulent eddies.  Nevertheless,
given the limited dynamical range available in these simulations, it would
not be surprising if the scale of the magnetic field has a dramatic impact
on the intermittency statistics.
   It is not clear how scale-dependent anisotropy changes the intermittency
   model
   in equation (\ref{mhdint}) and
   we will not discuss this issue further.
   Instead, we simply stress that we have found a striking 
   similarity between ordinary hydrodynamic turbulence and MHD turbulence
   in perpendicular directions. MG01 attributes the deviation of their spectrum
   from the Kolmogorov-type to the turbulence intermittency present in the
   MHD case. Since we do not reproduce their power spectrum, the fact that
our intermittency statistics do not support this conjecture is unsurprising. 
Clearly more studies of the issue are necessary.

\begin{figure}
\plottwo{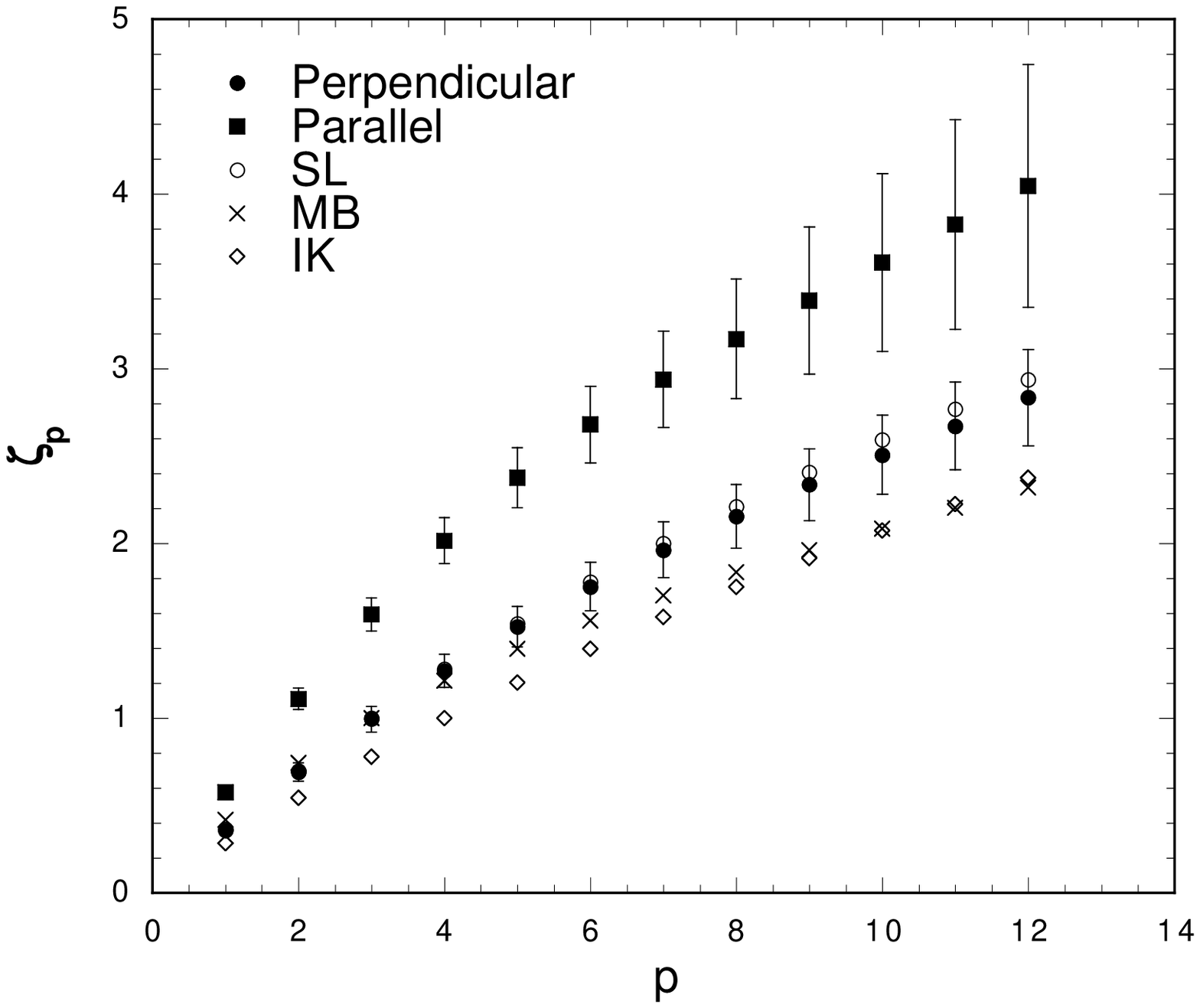}{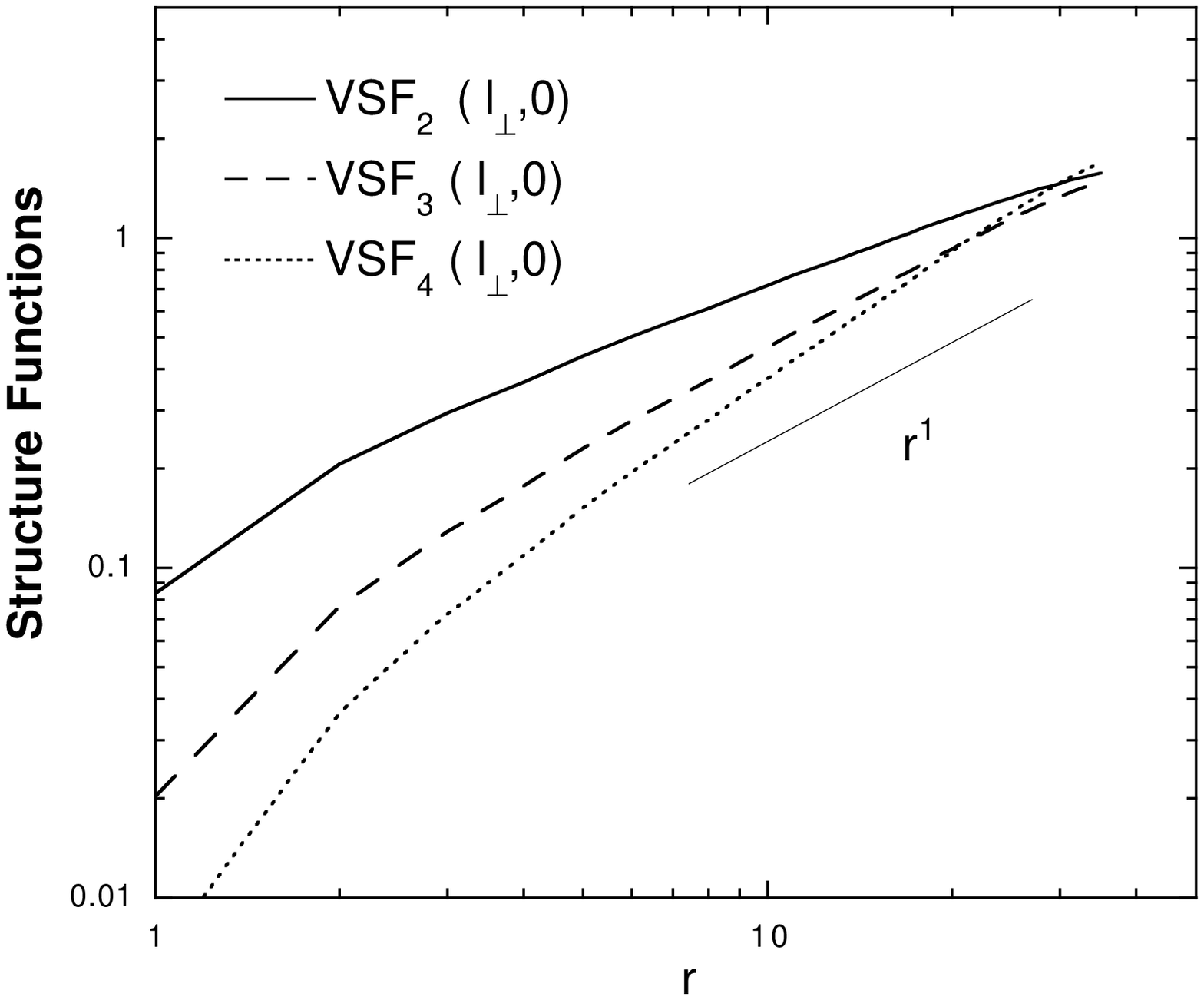}
\caption{  (a) Intermittency. Our result 
              (denoted by the filled circles) suggests
                       that MHD turbulence looks like
            ordinary hydrodynamic turbulence when viewed
            across the local field lines.
            SL represents the original She-Leveque model for
            ordinary hydrodynamic turbulence.
            IK and MB stand for the IK theory and the M\"uller-Biskamp model
            respectively.
            Error bars are for 1-$\sigma$ level.
            Run 144H-B$_0$1.   
         (b) The second, third, and fifth order longitudinal
          velocity structure functions.
          These are structure functions averaged over 
          the time interval (75,120).}

\end{figure}

In figure 4a, we also plot the scaling exponents (represented by filled 
squares) of 
longitudinal velocity structure functions {\it along} directions of
the local mean magnetic field.
Although we show only the exponents of longitudinal structure functions,
those of transverse structure functions follow a similar scaling law.
It is evident that intermittency along the local mean magnetic field directions
is completely different from that of previous (isotropic) models.
Roughly speaking, the scaling exponents along the directions of local 
magnetic field are 1.5 times larger than those of perpendicular
directions. 
Interestingly this result implies anisotropy becomes scale independent
under the following transformation: $(r_{\perp}, r_{\|})\rightarrow
(r_{\perp}, r_{\|}^{2/3})$.
This is consistent with the idea that eddies are stretched along the
directions of
the local mean magnetic field - if we shrink them in the scale-dependent 
manner described above along the local field lines,
the result is similar to ordinary hydrodynamic turbulence.
In this interpretation it is not surprising that MHD turbulence
looks similar to ordinary hydrodynamic turbulence across the
local mean magnetic field lines - the scaling relation in perpendicular directions
is not affected by the local mean magnetic field.
Clearly this result is hard to explain using previous models, 
for example the IK theory.
The error bars are larger for parallel directions because
fewer number of pairs are available for calculation of the structure functions
in these directions than perpendicular directions.

In Figure 4b, we plot average longitudinal velocity structure functions.
The slope of the third order structure function is very close to 1.
The third order structure function is slightly different from the
one discussed earlier in that we calculate $ < |\delta v_L|^3({\bf r}) >$
instead of  $ < (\delta v_L)^3({\bf r}) >$.

The second order exponent $\zeta_2$ is related to the the 1-D energy spectra:
$E(k_{\perp})\propto k_{\perp}^{-(1+\zeta_2)}$.
      Previous 2-D driven MHD calculations for $B_0=0$ by 
      Politano, Pouquet, \& 
     Carbone (1998) also found  $\zeta_2\sim 0.7$.
     However,  Biskamp, \& Schwarz (2001) obtained $\zeta_2\sim 0.5$
     from decaying 2-D MHD calculations with $B_0=0$.
Our result suggests that $\zeta_2$ is closer to $2/3$, rather than to $1/2$.
(It is not clear whether or not the scaling exponents follow the original S-L
model exactly. At the same time, our calculation shows that the original
S-L model can be a good approximation for our scaling exponents.
The S-L model predicts that $\zeta_2 \sim 0.696$.)
Therefore, our result supports the scaling law
   $E(k_{\perp}) \propto k_{\perp}^{-5/3}$ at least for velocity.
For the parallel directions, the results support 
$E(k_{\|}) \propto k_{\|}^{-2}$ although the uncertainty is large.


\section{The MHD Fluctuation Tensor}

For many purposes, e.g. cosmic ray propagation and acceleration, heat transfer
etc.,
it is necessary to know the tensor describing the statistics of magnetic
and velocity field. For those applications the one-dimensional 
spectrum described in
MG01 is not adequate and a more detailed description is necessary.

General second-rank correlation tensors are important tools
in the statistical description of turbulence.
Oughton, Radler, \& Matthaeus (1997) gave
a comprehensive formalism for the tensors for MHD turbulence
and we use their results as a starting point of our
argument.
Consider the velocity correlation tensor
\be
   R^v_{ij} = < v_i({\bf x})  v_j({\bf x}+{\bf r})>,
\ee
where the angle brackets denotes an appropriate ensemble average.
The Fourier transform of this tensor is
\be
  S^v_{ij} = < \hat{v}_i ({\bf k}) \hat{v}_j^* ({\bf k}) >,
\ee
where the asterisk denotes complex conjugate.
We can rewrite equation (20) of Oughton et al. (1997) as
\be
S_{ij}^v =
  \left[ \delta_{ij}-\frac{ k_ik_j }{ k^2 }\right] E^v({\bf k})
 +\left[ [e_ik_j+e_jk_i]({\bf e}\cdot{\bf k}) - e_ie_jk^2
     - \frac{ k_ik_j }{ k^2 } ({\bf e}\cdot{\bf k})^2 \right] F^v({\bf k})
 +X_{ij},
 \label{sij}
\ee
where $E^v({\bf k})$ is (3D) kinetic energy spectrum of all 
(shear + pseudo) modes,
$F^v({\bf k})$ is the difference of shear-Alfv\'en energy
and pseudo-Alfv\'en energy
at wave vector ${\bf k}$
divided by $k_{\perp}^2$, ${\bf e}$ is a unit vector along ${\bf B}_0$, and
$X_{ij}=
        -i[  \delta_{i\mu}\epsilon_{j\alpha \beta}
           + \delta_{j\mu}\epsilon_{i\alpha \beta}]e_{\alpha} k_{\beta}
           (e_{\mu} k^2 - k_{\mu} {\bf e}\cdot {\bf k})C^v
         + i\epsilon_{ij\alpha}k_{\alpha}H^v$
is a term that describes deviation from mirror symmetry.
In this paper we consider axisymmetric turbulence (caused by ${\bf B}_0$)
with mirror symmetry so that $X_{ij}\equiv 0$. 
We need only two scalar generating functions,
$E^v$ and $F^v$,
for the correlation tensor. 
This is consistent with Chandrasekhar (1951; see also Oughton
et al. 1997).

In this subsection, we will show that
the tensor is suitably described by
\be
 S_{ij} = A_1  \left[ \delta_{ij}-\frac{ k_ik_j }{ k^2 }\right]
 k_{\perp}^{-10/3} \exp\left( -A_2 \frac{ k_{\|} }{ k_{\perp}^{2/3} } \right),
\ee
where $A_1\sim B_0^2/L^{1/3}$ and $A_2\sim L^{1/3}$
are parameters.

First, we choose ${\bf e}$=(0,0,1), the direction of ${\bf B}_0$.
Then,
equation (\ref{sij}) becomes
\be
  S_{ij}^v =
  \left[ \delta_{ij}-\frac{ k_ik_j }{ k^2 }\right] E^v({\bf k})
 +\left[ [\delta_{3i}k_j+\delta_{3j}k_i] k_3
 - \delta_{3i}\delta_{3j}k^2
     - \frac{ k_ik_j }{ k^2 } k_3^2 \right] F^v({\bf k}).
\ee

In the absence of anomalous damping of the pseudo-Alfv\'en modes,
as in our simulations,
we can show that $F$ in above expression is negligibly small.
Note that $F= (S-P)/k_{\perp}^2$, where $S$ and $P$ are the squares of the
amplitudes of the shear- and pseudo-Alfv\'en modes (i.e. 3D energy spectra).
To evaluate $S$ and $P$, we measured their strength in global frame.
(It is non-trivial to correctly define
Alfv\'en modes and pseudo-Alfv\'en modes in the local frame.)
{}Figure 5 shows that they have similar strengths.
We assume that the same relation holds true
in the local frame.
Since $F$ is the difference between $S$ and $P$, it follows that
$F({\bf k})$ is small compared with $E({\bf k})$.

In the previous paragraph, we assumed that
there is no special damping mechanism for the pseudo-Alfv\'en modes.
However, it is known that pseudo-Alfv\'en modes in the ISM are
subject to strong damping due to free streaming of collisionless
particles along the field lines (Barnes 1966; Minter \& Spangler 1997).
When the pseudo-Alfv\'en modes are absent, equation (\ref{sij}) becomes
\be
  S_{ij}^v =
  \left[ \delta_{ij}-\frac{ k_ik_j }{ k^2 }\right] E^v({\bf k})
 +\left[ [\delta_{3i}k_j+\delta_{3j}k_i] k_3
 - \delta_{3i}\delta_{3j}k^2
 - \frac{ k_ik_j }{ k^2 } k_3^2 \right]\frac{ E^v({\bf k}) }{ k_{\perp}^2 }.
\ee
For $i,j=1,2$, this becomes $S_{ij} = (\delta_{ij} - k_ik_j/k^2) E
  - ( k_ik_j/k^2) ( k_{\|}/k_{\perp} )^2 E \approx
(\delta_{ij} - k_ik_j/k^2) E$.
And, it is easy to show that $S_{i3}=S_{3i}=0.$
This is easily understood when we note that
shear-Alfv\'en waves do not have fluctuations along ${\bf B}_0$.

In summary, the tensor reduces to
\be  \left( \begin{array}{ccc}
      (1- k_1^2/k^2) E &  -(k_1k_2/k^2 )E & -(k_1k_3/k^2) E \\
      -(k_1k_2/k^2 )E  &   (1- k_2^2/k^2) E & -(k_2k_3/k^2) E \\
       -(k_1k_3/k^2) E  & -(k_2k_3/k^2) E   &  (1- k_3^2/k^2)E \\
           \end{array}
    \right) \label{eq16}
\ee
for turbulence with both  Alfv\'enic and pseudo-Alfv\'enic components, and
\be  \left( \begin{array}{ccc}
      (1- k_1^2/k^2) E &  -(k_1k_2/k^2 )E &  0 \\
      -(k_1k_2/k^2 )E  &   (1- k_2^2/k^2) E & 0 \\
      0                &     0              &  0 \\
           \end{array}
    \right) \label{eq17}
\ee
for shear Alfv\'enic turbulence. In equation (\ref{eq17}), $E$ stands for
the energy of Alfven components only, which is roughly
one half of the $E$ in equation (\ref{eq16}).

\begin{figure}
\plottwo{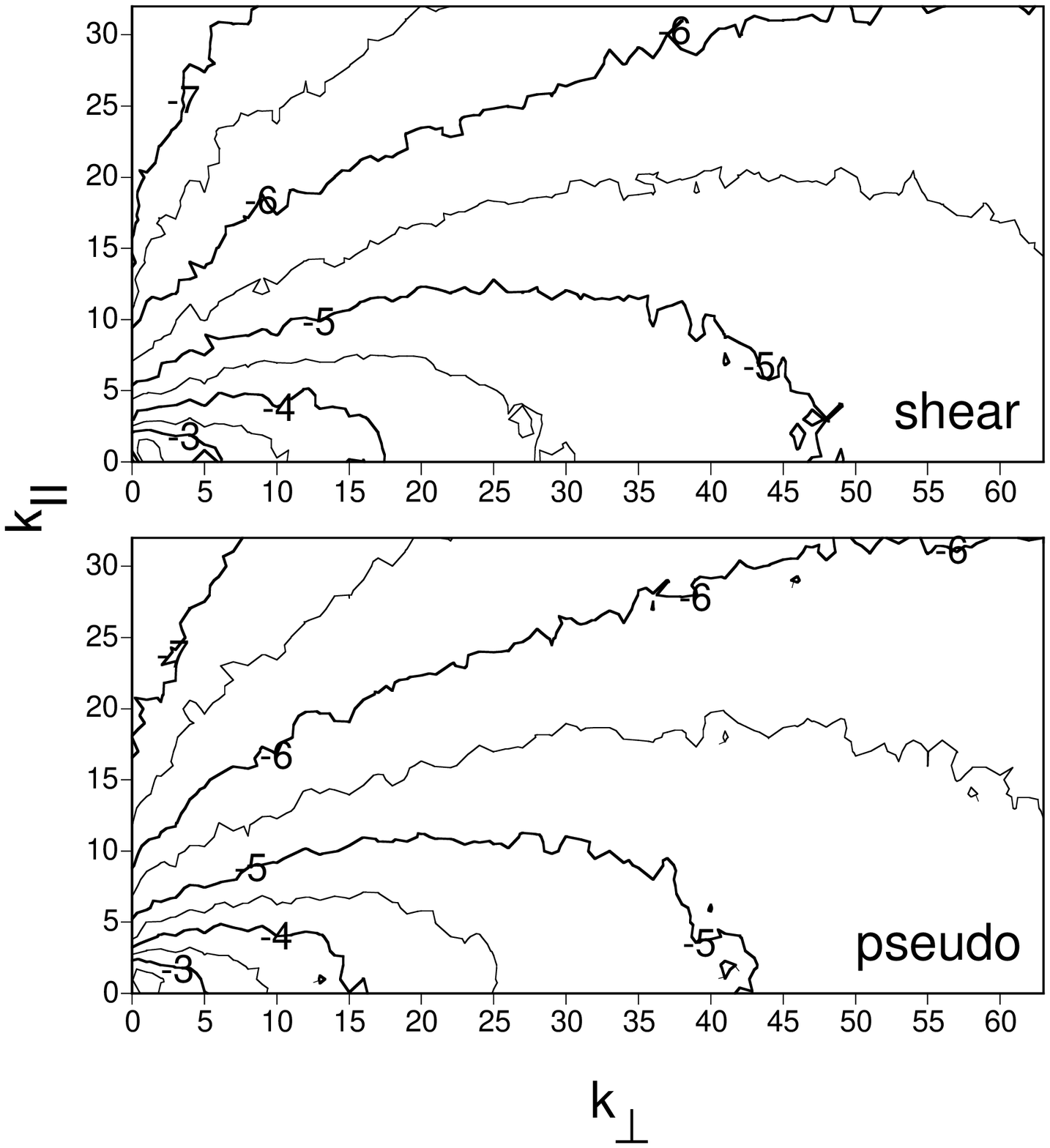}{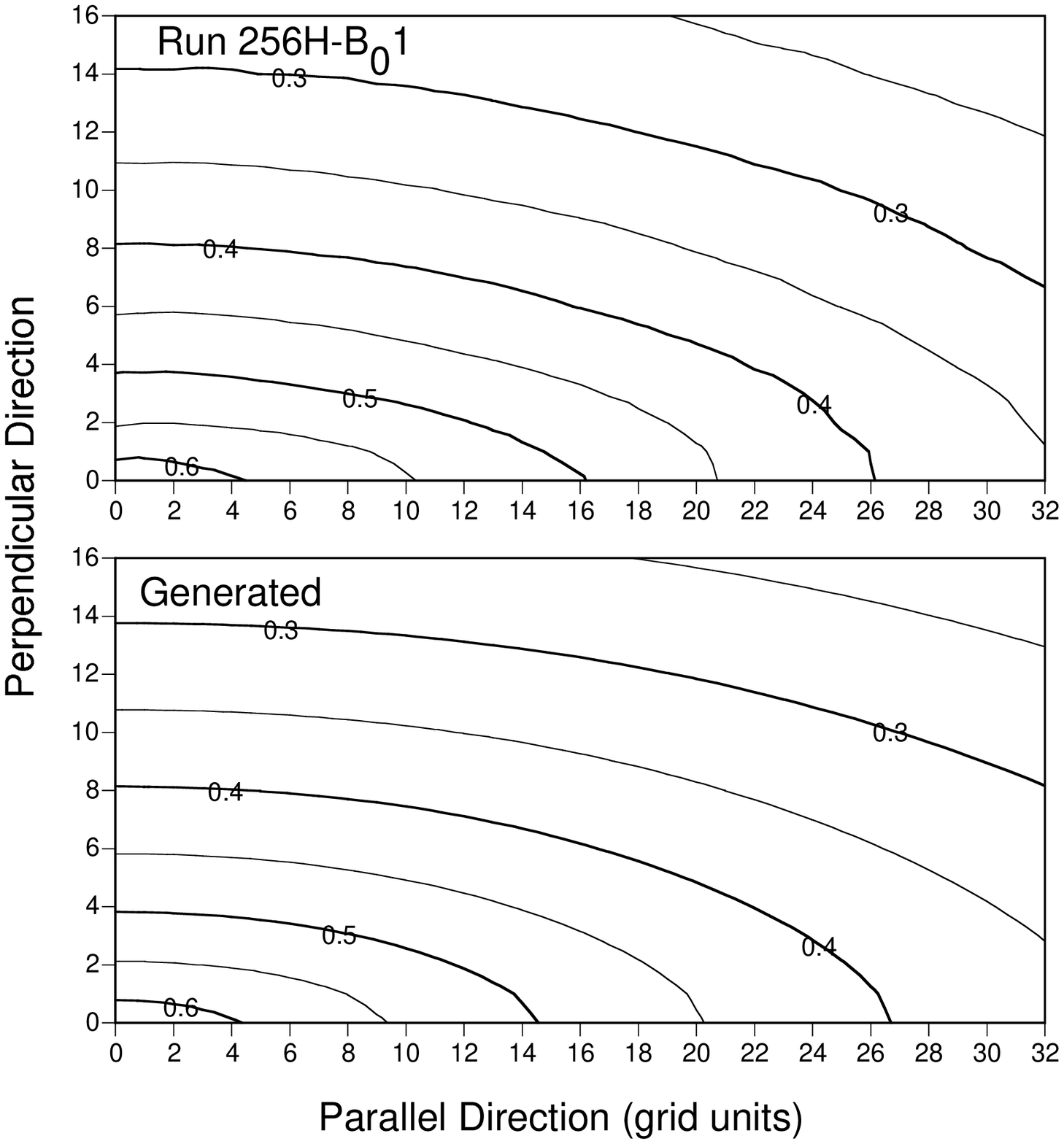}

\caption{   Energy distribution of shear- and
          pseudo-Alfv\'en waves.
   They have similar energy distributions in Fourier space (global frame).
   Run 256H-$B_0$1. Numbers by the contours are $\log_{10}(E_3({\bf k}))$.}

\caption{      Upper panel: velocity correlation function
                         from simulations.
                Lower panel: velocity correlation function generated using
                       the tensor in equation (\ref{tensor}).
           Parallel and perpendicular directions are taken with respect to
           local mean magnetic field. }

\end{figure}

The remaining issue is the form of $E$.
Note that the trace of $S_{ij}^v$ is $2E^v$.
In real space, the trace is the velocity correlation function.
Consequently, we can obtain $E^v$ through a FFT of the real-space velocity
correlation function, which is directly available from our data cube.
However, the velocity correlation function in real space contains
considerable numerical noise.  In order to minimize its effects while
obtaining an empirically useful form for $E$ we first
guess $E^v$ in Fourier space, do the FFT transform, and then
compare the transformed result with the actual velocity correlation function.
Since the trace of $S_{ij}^v$ is the (3-dimensional) energy spectrum
in Fourier space,
we start with the original expression in Goldreich \& Sridhar (1995)
given by equation (\ref{eq10}):
\be
  E_3(k_{\perp},k_{\|}) \sim \frac{ B_0^2 }{ k_{\perp}^{10/3}L^{1/3} }
             g(L^{1/3}\frac{ k_{\|} }{ k_{\perp}^{2/3} }),
\ee
where the functional form of $g(y)$ was not specified.
We have tried several functional forms for $g$ ; Gaussian
($g\propto \exp( -B  k_{\|}^2 / k_{\perp}^{4/3} )$),
exponential
($\exp( -B  k_{\|} / k_{\perp}^{2/3} )$),
and a step function.
We have found that an exponential form for $g$ gives
the best result (Figure 6).
{}Fig 6a is the actual data we obtained from our simulation and
{}Fig 6b is the Fourier-transformed velocity correlation function.
Note the similarity of the contours in both plots.
We conclude that
the tensor can be suitably described by equation
(\ref{eq16}) or (\ref{eq17}) with
\be
E(k_{\perp}, k_{\|})=(B_0/L^{1/3}) k_{\perp}^{-10/3}\exp\left(-L^{1/3}
     \frac{ k_{\|} }{ k_{\perp}^{2/3} } \right),
\label{tensor}
\ee
where $k_{\perp}$ and $k_{\|}$ should both be interpreted as the
absolute magnitudes of those wavevector components.

However, it is worth noting a clear limitation of 
equation (\ref{tensor}):
it has a discontinuous derivative near $k_{\|}=0$.
One way to overcome this difficulty is to use the Castaing function
(Castaing, Gagne \& Hopfinger 1990)\footnote{
     Our motivation for introducing the Castaing function is phenomenological,
not theoretical.  The theory of the distribution function for MHD turbulence is
uncertain, and far beyond the scope of this paper.}
\be 
 \Pi_{\lambda}(u) = \frac{1}{2\pi \lambda}
   \int_{0}^{\infty} \exp \left(-\frac{u^2}{2\sigma^2}\right)
    \exp \left( -\frac{  \ln^2(\sigma/\sigma_0) }{ 2\lambda^2 } \right)
    \frac{ d\sigma }{ \sigma^2 },
\ee
which is smooth near zero 
but looks exponential over a broad range.
It is possible to see that for $\lambda = 1$ 
and $\sigma_0 = k_{\perp}^{2/3}/L^{1/3}$,
\be
\exp\left(-L^{1/3}
     \frac{ k_{\|} }{ k_{\perp}^{2/3} } \right)
 \approx 
 \frac{1}{2\pi }
   \int_{0}^{\infty} \exp \left(-\frac{k_{\|}^2}{2\sigma^2}\right)
    \exp \left( -\frac{  \ln^2(L^{1/3} \sigma/ k_{\perp}^{2/3}   ) }
                      { 2 }                                      \right)
    \frac{ d\sigma }{ \sigma^2 }
\ee
However, for many practical applications, we feel that 
the expression in equation (\ref{tensor})
is adequate. 
For instance, in a forthcoming paper (Yan et al. 2001)
this tensor is used for describing cosmic ray
propagation and we find a strong suppression of cosmic ray 
scattering compared with
the generally accepted estimates (for example, Schlickeiser 1994).
However, if the behavior around $k_{\|}$ is important, the Castaing function
would be preferred.  In our simulations there is no way to distinguish between 
exponential and Castaing distributions.


\section{Decay of MHD Turbulence}

Turbulence plays a critical role in molecular cloud support and star 
formation and the issue of the time scale of turbulent decay is vital for
understanding these processes.

If MHD turbulence decays quickly then serious
problems face researchers attempting to explain important observational 
facts, i.e.  turbulent  motions seen within molecular clouds without
star formation (see Myers 1999) and rates of star formation (Mckee 2000).
Earlier studies attributed the rapid decay of turbulence to compressibility
effects (Mac Low 1999). Our present study, as well as earlier ones  
(CV00, MG01), shows that turbulence decays rapidly even
in the incompressible limit. This can be understood in the framework of
GS95 model where mixing motions perpendicular to magnetic field lines
form hydrodynamic-type eddies. Such eddies, as in
hydrodynamic turbulence, decay in one eddy turnover time.

How grave is this problem?  Some possibilities
for reconciling theory with observations were studied earlier.
For instance, some problems
may be alleviated if the injection of energy happens on the large
scale, the eddies are huge, and the corresponding time scales, are
much longer (see Lazarian 1999). 
The fact that the turbulence decays according to a power-law, rather
than exponentially, also helps. Indeed, if turbulent energy decays
as $t^{-1}$, as suggested by Mac Low wt al. (1998), a substantial level of
turbulence should persist after 4-5 turnover times.

There is, however, another property of astrophysical turbulence related
to the peculiar nature of the energy injection. It is accepted that
sources of interstellar turbulence are localized. As a
result, there is a substantial imbalance between the ingoing and
outgoing energy flux surrounding every source.
Below we consider the effect of this 
imbalance on the turbulence decay time scale.

For an imbalanced turbulence,
it is useful to consider the Elsasser variables, ${\bf z}^{\pm}
={\bf v} \pm {\bf b}$, which
describe wave packets traveling in opposite directions along
the magnetic field lines.
Imbalanced turbulence means that wave packets traveling in one
direction (say, ${\bf z}^{+}$)
have significantly larger amplitudes than the other.
In astronomy, many energy sources are localized.
For example, SN explosions and OB winds are typical point energy sources.
Furthermore, astrophysical jets from YSOs are believed to be
highly collimated.
With these localized energy sources, it is natural to think
that interstellar turbulence is typically  imbalanced.
In fact, the concept of an imbalanced cascade is not new.
Earlier papers (e.g., Matthaeus, Goldstein \& Montgomery 1983;
Ting, Matthaeus \& Montgomery 1986; Ghosh, Matthaeus \& Montgomery 1988)
have addressed
the role and evolution of cross-helicity ($\equiv {\bf v}\cdot{\bf b}$).
Since $ 4<{\bf v}\cdot{\bf b}> = <({\bf z}^+)^2> -<({\bf z}^-)^2>$,
non-zero cross-helicity implies an imbalanced turbulent cascade.
These works, however, were mainly concerned with the growth of
imbalance in decaying turbulence.
Ghosh et al. (1988) investigated the evolution of cross-helicity and
various spectra
in driven turbulence.
Hossain et al. (1995) discussed the effects of cross helicity and energy
difference $D=<v^2>-<b^2>$ on the decay of turbulence. Their
low resolution 3D numerical simulations show the effect of
cross helicity, although that effect is not very conspicuous.
A further study of imbalanced turbulence was given in MG01, who
also suggested a connection between spontaneous appearance of
local imbalance in the turbulent cascade and intermittency in
MHD turbulence.

In this subsection, we explicitly relate the degree of imbalance and
the decay time scale of turbulence in the presence of a strong
uniform background field. 

\begin{figure}
\plottwo{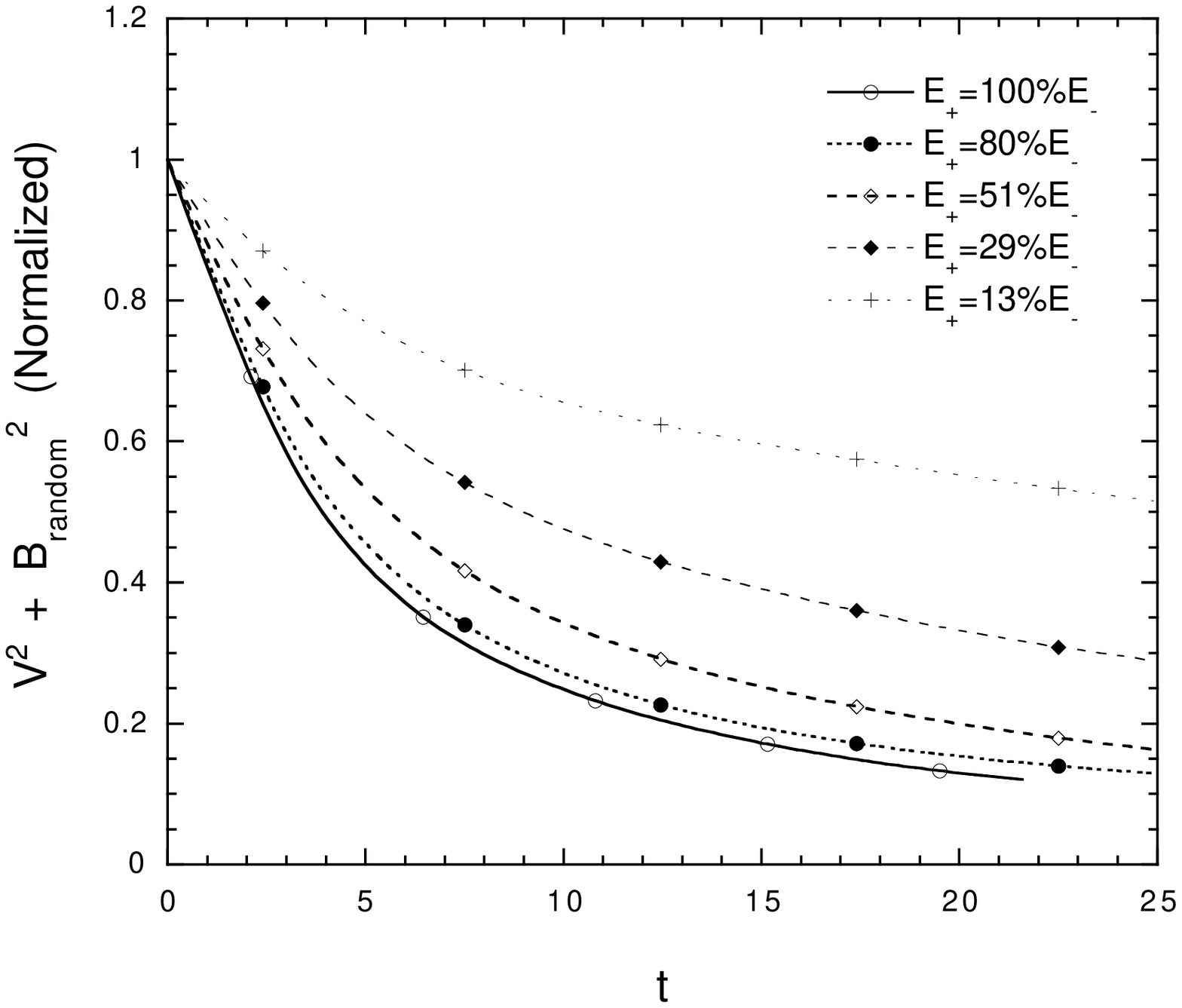}{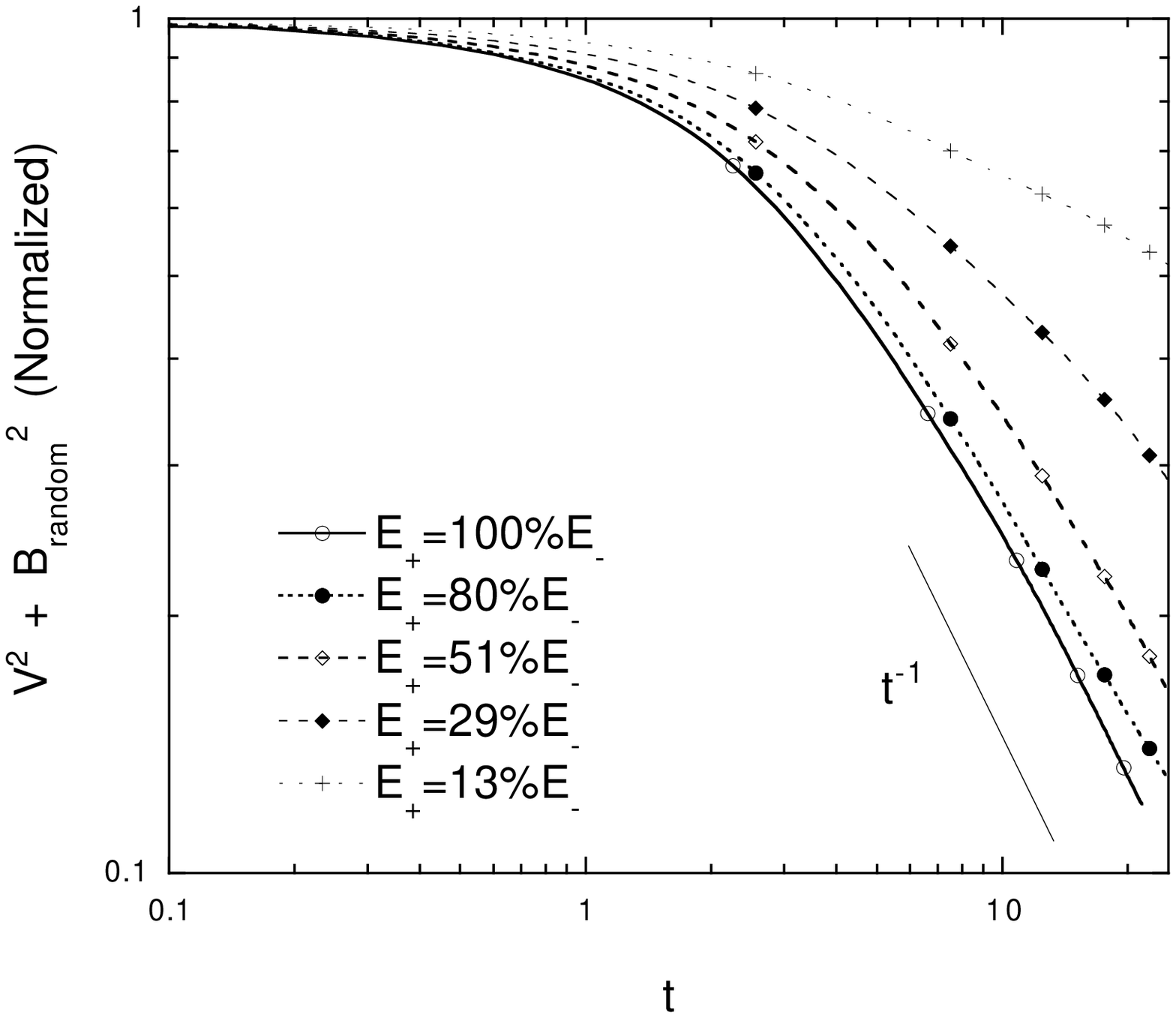}
\caption{  (a) Decay of unbalanced turbulence.
                 Imbalanced cascade can extend
                decay time.
            It is clear that the decay of turbulence depends 
      on the degree of imbalance. Run 144H-$B_0$1.     
    (b) Same as (a). Log-log scale.  Note that about 3 time units is 
     one eddy turn over time.   }

\end{figure}

In Figure 7 we demonstrate that an imbalanced cascade does extend
the lifetime of MHD turbulence.
We use the run 144H-$B_0$1 to investigate the decay time scale.
We ran the simulation up to t=75 with non-zero driving forces.
Then at t=75, we turned off the driving forces
and let the turbulence decay.
At t=75, there is a slight imbalance between upward and downward moving
components ($E_+ = 0.499$ and $E_- = 0.40$).
This results from a natural fluctuation in the simulation.
The case of $(E_{+})_{t_0} = 80\% (E_{-})_{t_0}$ corresponds to
the simulation that starts off from this initial imbalance.
In other cases, we either increase or decrease the energy of ${\bf z}^{-}$
components and, by turning off the forcing terms, let the turbulence decay.
We can clearly observe that imbalanced turbulence extends
the decay time scale substantially.
Note that we normalized the initial energy to 1.
The y-axis is the total (=up $+$ down) energy.

In Figure 7b, we re-plot the Figure 7a in log-log scale.
{}For balanced case (i.e. zero cross-helicity case; solid curve), 
the energy decay follows a power law $E(t)\propto t^{\alpha}$, where
$\alpha$ is very close to 1. This result is consistent with 
previous 3D result by Hossain et al. (1995).
Note that hydrodynamic turbulence decays faster than this.
For example, Kolmogorov turbulence decays as $E(t)\propto t^{-10/7}$.
In this sense, it may not be absolutely correct to say that
both hydro and MHD turbulence decay within one eddy turnover time.
However, note that the power law
does not hold true from the beginning of decay.
We believe that, at the initial stage of decay, 
the speed of decay is still roughly
proportional to the large-scale eddy turn over rate.

How far does a wave packet travel when there is an imbalance?
Consider the equations governing an imbalanced cascade.
{}From the MHD equations, Hossain et al. (1995) and MG01 derived a simple
dynamical model for imbalanced turbulence.
{}For decaying turbulence, they found
\bea
 \frac{ d E_+ }{ dt } & = & - \frac{ E_+ E_-^{1/2} }{ L }, \\
 \frac{ d E_- }{ dt } & = & - \frac{ E_- E_+^{1/2} }{ L },
\eea
where $L$ is the largest energy containing eddy scale.\footnote{ For 
     our current purposes, these simple system of equations are
     enough. For more rigorous equations on the evolutions of $E_{\pm}(k)$,  
     readers may refer to earlier closure equations, e.g. 
     Grappin et al. (1982). See also Hossain et al. (1995) for
     time evolution of $L$. }
{}From these coupled equations, they showed that
imbalance grows exponentially in decaying turbulence.
Now let us consider a large amplitude  wave packet traveling
in an already (weakly) turbulent background medium.
Suppose that the large amplitude wave packet corresponds to ${\bf z}^+$.
Using the simplified equations,
we obtain
$\dot{E_+}/E_+ = - E_-^{1/2}/L$. If the background turbulence has a
constant amplitude, the  ${\bf z}^+$ wave decays exponentially.
It can travel
\be
 v^{+} \Delta t \sim (E_+^{1/2}/E_-^{1/2})L,
\ee
where we use $ v^{+}\sim E_+^{1/2}$.
This means that the wave packet can travel a long distance when imbalance is
large (i.e. $E_+^{1/2} \gg E_-^{1/2}$).
In real astrophysical situations, the problem is not as simple as this.
Instead, the wave packet and the background turbulence can have different
length scales (as opposed to the single scale $L$ in the equations).
We also need to consider the fact that the amplitude of the background
turbulence does not stay constant and the front of the wave packet decays
faster than the tail of the packet.  Finally, MHD turbulence can influence
the pressure support only if turbulent motions are at least comparable
to the sound speed, which obviously requires a fully compressible
treatment.  Preliminary calculations with
a compressible code (Cho \& Lazarian 2002) show marginal coupling
of compressible and incompressible motions, while the development of 
the parametric instability (Fukuda \& Tomoyuki 2000) requires more time
to develop. We plan to investigate these possibilities in future.

In this section, we found that turbulence decay time can be slow.
This finding is very important for many astrophysical problems.

\section{Discussion}

How relevant are our calculations for the ``big picture''? First of all,
they provide more support for the GS95 theory, indicating that for the
first time we have an adequate, if approximate, theory of MHD turbulence. 
Second,
they extend the theory by treating new cases, e.g. an imbalanced cascade.
Third, they establish new scaling relations and determine critical
parameters,
e.g. the functional form of $g$ in equation (\ref{eq10}), that will 
allow the theory to be applied to
various astrophysical circumstances.

Our calculations are made within an intentionally simplified model,
which is based on the physics of an incompressible fluid.
This surely raises the question of the applicability of our scaling 
relations and conclusions to realistic circumstances. 
There are situations where
our scalings should be applicable. For instance, turbulence at very small
scales is small-amplitude and therefore essentially incompressible. 
Processes that depend on the fine structure of turbulence, like
scintillation, reconnection, and the propagation of cosmic rays of moderate
energies should be well described using our results.

If we consider the interstellar medium at larger scales,
it is definitely compressible, has a whole range of energy
injection/dissipation
scales (see Scalo 1987), and the relative role of vortical versus compressible
motions being unclear. Nevertheless, we believe that our
simplified
treatment may still elucidate some of the basic processes.
To what extend
this claim can be justified will be clear when we 
compare compressible and incompressible results. However, if we accept
that fast and slow MHD modes are subjected to fast collisionless damping
(see Minter \& Spangler 1997) the remaining modes are 
incompressible Alfven modes.  Those should
be well described by our model when turbulence is supersonic but
sub-Alfvenic. Our preliminary results (Cho \& Lazarian 2002) show
that the coupling of the modes is marginal even in compressible
regime. 
Incidentally, recent studies of turbulence of HI in both
our Galaxy and SMC (Lazarian 1999,
Lazarian \& Pogosyan 2000, Stanimirovic \& Lazarian
2001) show the spectra of velocity and density consistent with the Kolmogorov
scalings\footnote{If density acts as a passive scalar its spectrum
mimics that of velocity over the inertial range.}.

Our approach is complementary to MG01. They studied turbulence
in the regime when the magnetic energy is substantially larger than
the kinetic energy at the energy injection scale. Physically their
regime reflects better the properties of turbulence on small scales,
where the magnetic energy is indeed dominant. In our calculations the
kinetic energy is equal to the magnetic energy at the energy injection scale
and therefore they reflect, for instance, what is happening in the
interstellar medium at the large scales. Our results show that even
on those scales GS scaling is applicable. This suggests that the
astrophysical turbulence may be well tested not only via scintillations, that
reflect properties of the turbulence on the small scales, but with
other techniques, e.g. synchrotron emission.


\section{Summary}

Our findings can be summarized as follows:

The energy cascade time scale at a length scale $l$ ($\sim 1/k$)
is proportional to $l^{2/3}$ ($k^{-2/3}$), which is consistent with
the prediction of the GS95 model and numerical simulations by
MG01 who used a different method to obtain
this scaling. In this respect MHD turbulence is similar to
hydrodynamic turbulence.
This scaling is distinctly different from the prediction of
Iroshnikov-Kraichnan theory, $t_{cas} \propto l^{1/2}$.

We found that velocity fluctuations in the direction parallel to local
magnetic field  follow a similar scaling for 
both Alfvenic and pseudo-Alfvenic modes. We determined that 
parallel motions due to pseudo-Alfven
perturbations  obey the following scaling: $v_{\|}\sim k_{\|}^{1/2}$.
This finding is important for practical applications, e.g. for description
of dust grain motion.

To study intermittency we calculated higher order longitudinal
velocity structure functions in directions perpendicular to
the local mean magnetic field and found that the scaling exponents
are close to
$\zeta_p^{SL}=p/9+2[1-(2/3)^{p/3}]$. As this coincides with
the She-Leveque model of intermittency in hydrodynamic flow
we speculate that there may be more similarities between
magnetized and unmagnetized turbulent flows than has been
previously anticipated.

We obtained correlation tensors
which provide a good fit for our numerical results. These tensors
are valuable for theoretical applications, e.g. to describe
cosmic ray transport.

We found that the rate at which MHD turbulence decays depends on the degree
of energy imbalance between wave packets traveling in opposite directions.
A substantial degree of imbalance can substantially 
extend the decay time scale of the MHD turbulence and the distance
the turbulence can propagate from the source.

\acknowledgements

The authors are thankful to Peter Goldreich for
attracting our attention to the case of imbalanced cascade and
for many valuable discussions.
We also thank John Mathis for many useful suggestions and 
discussions. AL and JC acknowledge the support of NSF Grant...., and
EV acknowledges the support of NSF Grant AST-0098615.
This work was partially supported by National Computational Science
Alliance under CTS980010N and AST000010N and
utilized the NCSA SGI/CRAY Origin2000.

\clearpage

\begin{deluxetable}{ccccl}
\tablecaption{Parameters}
\tablewidth{0pt}
\tablehead{
\colhead{Run \tablenotemark{a}} & \colhead{$N^3$} & \colhead{$\nu$} & \colhead{$\eta$} &
\colhead{$B_0$}
}
\startdata

144H-$B_0$1 & $144^3$ & $3.20 \times 10^{-28}$  &  $3.20 \times 10^{-28}$  & 1
\\

256H-$B_0$1 & $256^3$ & $6.42 \times 10^{-32}$  &  $6.42 \times 10^{-32}$  & 1
\\

\enddata

\tablenotetext{a}{ For $256^3$ (or $144^3$) grids,
we use the notation 256X-Y (or 144X-Y),
where X = H or P refers to hyper- or physical viscosity;
Y = $B_0$1 refers to the strength of the external magnetic fields.}

\end{deluxetable}


\begin{references}

\reference{} Barnes, A. 1966, Phys. Fluids, 9, 1483

\reference{} Bhattacharjee, A., Ng, C.S., \& Spangler, S.R. 1998, ApJ, 494, 409

\reference{} Biskamp, D., \& Schwarz, E. 2001, Phys. Plasmas, 8(7), 3282

\reference{} Borue, V. \& Orszag, S.A., 1996, J. Fluid Mech., 306, 293

\reference{} Castaing, B., Gagne, Y. \& Hopfinger, E., 1990, Physica D, 46, 177

\reference{} Cho, J. \& Lazarian, A. 2002, in preparation
\reference{} Cho, J., \& Vishniac, E. T. 2000a, ApJ, 539, 273 (CV00)
\reference{} Cho, J., \& Vishniac, E. T. 2000b, ApJ, 538, 217



\reference{} Chandrasekhar, S., 1951, Proc. R. Soc. London, Ser. A, 207, 301

\reference{} Galtier, S., Nazarenko, S., Newell, A., \& Pouquet, A. 2000,
             J. Plasma Phys., 63, 447
\reference{} Ghosh, S., Matthaeus, W.H. \& Montgomery, D.C.
             1988, Phys. Fluids, 31(8), 2171
\reference{} Goldreich, P. \& Sridhar, H. 1995, ApJ, 438, 763 (GS95)
\reference{} Goldreich, P. \& Sridhar, H. 1997, ApJ, 485, 680
\reference{} Grappin, R., Frisch, U., Pouquet, A. \& Leorat, J., 1982, 
                             A\&A, 105, 6
\reference{} Fukuda, N. \& Hanawa, T. 1999, ApJ, 517, 226

\reference{} Hossain, M., Gray, P.C., Pontius, D.H., \& Matthaeus, W.H. 1995,
                 Phys. Fluids, 7(11), 2886

\reference{} Iroshnikov, P., 1963, Astron. Zh., 40, 742
             (English version: 1964, Sov. Astron., 7, 566)
\reference{} Kolmogorov, A. 1941, Dokl. Akad. Nauk SSSR, 31, 538

\reference{} K\'ota, J. \& Jokipii, J. R., 2000, ApJ, 531, 1067
\reference{} Kraichnan, R., 1965,
             Phys. Fluids, 8, 1385

\reference{} Lazarian, A. 1995, A\&A, 293, 507

\reference{} Lazarian, A. 1999, in Plasma turbulence and energetic particles, 
  ed. M. Ostrowski \& R. Schlickeiser (Cracow), 28, astro-ph/0001001

\reference{} Lazarian, A. \& Pogosyan, D. 2000, ApJ, 537, 720L

\reference{} Lazarian, A.\& Vishniac, E. T. 1999, ApJ, 517, 700

\reference{} Leamon, R. J., Smith, C. W., Ness, N. F., \& Matthaeus, W. H. 
1998, J. Geophys. Res., 103, 4775

\reference{} McKee, C.F. 1999, astro-ph/9901370
\reference{} Maron, J. \& Goldreich, P. 2001, ApJ, 554, 1175 (MG01)

\reference{} Matthaeus, W.H., \& Goldstein, M. L. 1982, J. Geophys. Res.,
             87, 6011
\reference{} Matthaeus, W.H., Goldstein, M. L., \& Montgomery, D.C.
             1983, Phys. Rev. Lett., 51(16), 1484

\reference{} Matthaeus, W.H., Oughton, S., Ghosh, S. and Hossain, M.,
             1998, Phy. Rev. Lett. 81, 2056
\reference{} Mac Low, M., Klessen, R., \& Burkert, A. 1998, Phys. Rev. Lett., 
             80(13), 2754
\reference{} Mac Low, M. 1999, ApJ, 524, 169

\reference{} Milano, L.J., Matthaeus, W.H., Dmitruk, P. \& Montgomery, D.C., 
              2001, Phys. Plasmas, 8(6), 2673
\reference{} Minter, A. \& Spangler, S. 1997, ApJ, 485, 182

\reference{} Montgomery, D.C. 1982, Physica Scripta, T2/1, 83

\reference{} Montgomery, D.C. \& Matthaeus, W.H., 1995, ApJ, 447, 706

\reference{} Montgomery, D.C. \& Turner, L., 1981, Phys. Fluids, 24(5), 825

\reference{} M\"uller, W.-C. \& Biskamp, D. 2000, Phys. Rev. Lett., 84(3) 475

\reference{} Myers, P.C., 1999, in
                   The Origin of Stars and Planetary Systems,
                   ed. by Charles J. L. and Nikolaos D.K.,
                   Kluwer Academic Publishers, 1999, p.67


\reference{} Ng, C.S., \& Bhattacharjee, A. 1996, ApJ, 465, 845
\reference{} Politano, H. \& Pouquet, A., 1995,
             Phys. Rev. E, Vol. 52, No.1, 636
\reference{} Politano, H., Pouquet, A., \& Carbone, V. 1995, 
                 Europhys. Lett., 43(5), 516
\reference{} Politano, H., Pouquet, A. \& Sulem, P.L. 1995, Phys. Plasmas,
             2(8), 2931

\reference{} Rosenbluth, M.N., Monticello, D.A., Strauss, H.R., \& White, R.B.,
             1976, Phys. Fluids, 19, 1987

\reference{} Scalo, J. M. 1987, in Interstellar Processes,
             ed. D. J. Hollenbach \& H. A. Thronson (Dordrecht: Reidel), 349

\reference{} Schlickeiser, R., 1994, ApJS, 90, 929

\reference{} She, Z.-S. \& Leveque, E. 1994, Phys. Rev. Lett., 72(3), 336 (S-L)

\reference{} Shebalin J.V., Matthaeus, W.H. \& Montgomery, D.C., 1983,
             J. Plasma Phys., 29, 525

\reference{} Sridhar, H., \&  Goldreich, P., 1994, ApJ, 432, 612

\reference{} Stanimirovic, S., \& Lazarian, A. 2001, ApJ, in press

\reference{} Strauss, H. R. 1976, Phys. Fluids, 19, 134

\reference{} Ting, A., Matthaeus, W.H. \& Montgomery, D.C. 1986,
             Phys. Fluids, 29(10), 3261
\reference{} Yan, H., Lazarian, A. \& Cho, J. 2001, in preparation
\reference{} Yan, H., Lazarian, A. \& Zweibel, E. 2001, in preparation

\reference{} Zank, G. P., \& Matthaeus, W. H. 1992, J. Plasma Phys., 48, 85
\end{references}
\end{document}